\documentclass[11pt,letterpaper]{article}
\usepackage[utf8]{inputenc}
\usepackage{paralist}
\usepackage{multicol}
\usepackage{multirow}
\usepackage{float}
\usepackage{amsmath}
\usepackage{amsfonts}
\usepackage{amssymb}
\usepackage{graphicx}
\usepackage{geometry}
\usepackage[round]{natbib}
\usepackage{subfig}
\usepackage{tabularx}
\usepackage{tabulary}
\usepackage[table]{xcolor}
\usepackage{caption}
\usepackage{physics}
\usepackage{hyperref}
\usepackage[parfill]{parskip}
\usepackage{cleveref}

\hypersetup{
	bookmarksopen=true,
	colorlinks=true,
	linkcolor=blue,
	citecolor=blue,
	urlcolor=blue,
	pdftitle={General purpose user element routine for bloch analysis in FE codes},
	pdfauthor={C. Valencia, J. Gomez, N. Guarin-Zapata},
	pdfkeywords={Periodic materials, Dispersion relation, Bloch analysis,  User element routines},
	pdfsubject={Bloch analysis},
	pdfpagemode=UseOutlines,
	pdfstartview=FitH
}

\title{\textbf{A general purpose element-based approach to compute dispersion 
relations in periodic materials with existing finite element 
codes}\footnote{Preprint of an article published in Journal of Theoretical and 
Computational Acoustics, © 
\href{https://www.worldscientific.com/worldscinet/jca}{World 
Scientific  Publishing Company}}}

\author{Camilo Valencia\footnote{Corresponding author: cvalen20@eafit.edu.co}, 
Juan Gomez and Nicolás Guarín-Zapata\\ 
\\
Universidad EAFIT \\ Departamento de Ingeniería Civil \\ Medellín, Colombia}                 
\date{\today}

\begin{document}

\maketitle

\begin{abstract}
In most of standard Finite Element (FE) codes it is not easy to calculate dispersion relations from periodic materials. Here we propose a new strategy to calculate such dispersion relations with available FE codes using user element subroutines. Typically, the Bloch boundary conditions are applied to the global assembled matrices of the structure through a transformation matrix or row-and-column operations. Such a process, is difficult to implement in standard FE codes since the user does not have access to the global matrices. In this work, we apply those Bloch boundary conditions directly at the elemental level. The proposed strategy can be easily implemented in any FE code. This strategy can be used either in real or complex algebra solvers. It is general enough to permit any spatial dimension and physical phenomena involving periodic structures. A detailed process of calculation and assembly of the elemental matrices is shown. We verify our method with available analytical solutions and external numerical results, using different material models and unit cell geometries.\\


\end{abstract}

\section*{Introduction}
Periodic materials either in the form of composites, like in the so-called 
phononic crystals, or those in the more exotic family of metamaterials, have 
received much attention during the recent years thanks to their various 
attractive features across different disciplines 
\citep{banerjee2011introduction}. These artificial materials are designed to 
meet specific functionalities through modifications at the microstructural 
level thus allowing effective macroscopic responses non-present in nature. 
Among the most interesting responses one finds negative mass density, negative 
refraction, and electromagnetic cloaking 
\citep{hussein2014,goldsberry2018,norris2012}, while particular applications 
are identified in the works of \cite{hiett2002application} in photonic 
crystals, \cite{porter2003scattered} in micro-structured soils and 
\cite{michel1999effective} in composite materials. A particular aspect of the 
response of periodic media, present within different physical contexts, is the 
existence of bandgaps or specific frequency ranges where wave propagation is 
forbidden. Such finite frequency gaps are known to be intimately related or 
equivalent to the dispersive nature of these micro-structured materials where 
wave propagation velocity depends upon frequency. Subsequently , a key step in 
the conception of a periodic material within the context of wave propagation 
response is the micro-structural design of the fundamental material cell in 
such a way that it delivers bandgaps at specific frequency ranges and with 
particular properties. This characterization of the response is achieved 
typically in terms of the dispersion diagram or band structure of the material. 
Mathematically, and more interestingly numerically, the band structure for a 
periodic material can be obtained via Bloch's theorem, 
\citep{bloch1929quantenmechanik}, which takes advantage of the fact that the 
material is composed of a single fundamental cell distributed periodically 
throughout space. This work deals with the numerical determination of the band 
structure through the finite element method within different physical settings.

There are several issues that must be solved when computing band diagrams or dispersion relationships for periodic media with commercially available finite element codes. First, the wide majority of general purpose finite element packages available in the market are not equipped with the intrinsic functionality required to conduct the so-called Bloch analysis. Secondly, the imposition of Bloch periodic boundary condition (henceforth referred to like BBCs), implies the modification of global coefficient matrices by transformation arrays \citep{mcgrath1994phased} which perform row-column operations \citep{mcgrath1996periodic}. Such access and manipulation of the global coefficient matrix, is not only cumbersome but also prohibited in most commercial packages. Here we propose a novel, yet simple technique to compute dispersion relations in periodic materials with commercial finite element codes with intrinsic user-element functionalities. The main distinguishing, and most appealing feature in our method is the fact that all manipulations are conducted directly at the element level thus allowing for the treatment of a wide variety of physical contexts and problem dimensions. The proposed approach is motivated by the work from \cite{pask2005} who used a similar technique to solve one dimensional problems after transforming the differential equation not to consider Bloch's theorem but the simpler case of periodic boundary conditions. That idea was later extended by \cite{sukumar2009} who incorporated BBCs although in problems restricted to scalar fields. The method proposed in this work extends these ideas but it is valid for both, periodic and Bloch periodic boundary conditions, in addition to the possibility of considering general problems regardless of their dimensionality and kinematic assumptions. Thus it is valid for a wide variety of scalar and vector valued functions belonging to different physical contexts. Furthermore, the numerical consideration of Bloch's theorem in periodic media requires the solution of a complex-valued eigenvalue problem which can be solved using a 2-mesh approach as proposed by \cite{aaberg1997usage} or through the direct implementation of the numerical scheme in a complex-algebra-based finite element code \citep{langlet1995analysis}. Our element-based procedure can be implemented in both formats.

In addition to this introduction, the paper contains four more sections. Section 1 presents, for completeness, some theoretical background related to periodic materials and the finite element formulation of Bloch analysis. Section 2, formally introduces the proposed strategy to apply BBCs directly into the local elemental-based arrays. In the subsequent section we conduct several verification exercises intended to show the generality of the numerical tool. Particularly,  we solved elastodynamic problems for classical and Cosserat or micropolar based materials. This last cases is intended to show the possibility of implementing different kinematic assumptions. We considered, scalar problems corresponding to out-of-plane waves and vector-valued problems for in-plane waves. Additionally we also conducted numerical exercises to test the sensitivity of the implementation to the cell-size. In the different verification problems considered we used analytic and numerical results reported in the literature. As a complement we also included a package of supplementary material consisting of: (i) a fully functional version of FEAPpv to calculate the dispersion relations from a periodic material and (ii) a document explaining the usage of this version of FEAPpv with an example of a homogeneous material unit cell.

\section{Wave propagation in periodic media}
\begin{figure} [h]
\centering
\subfloat[]{\includegraphics[width=0.22\textwidth]{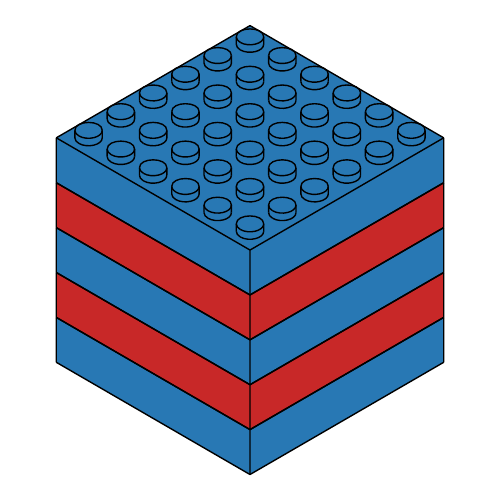}}
\subfloat[]{\includegraphics[width=0.22\textwidth]{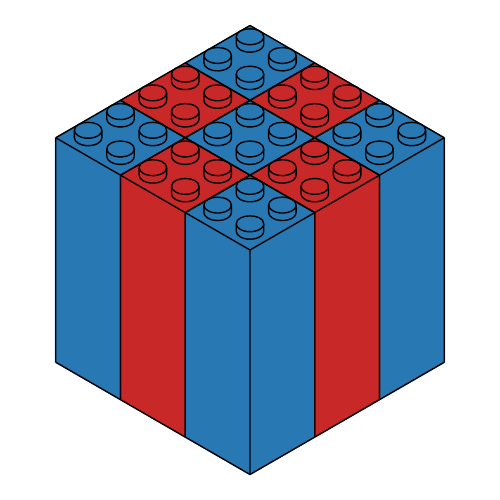}}
\subfloat[]{\includegraphics[width=0.22\textwidth]{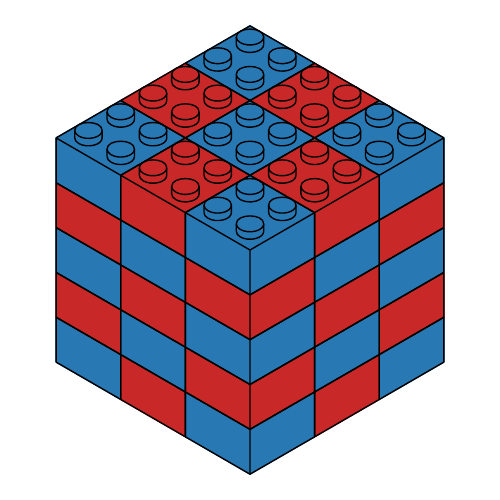}}
\subfloat[]{\includegraphics[width=0.32\textwidth]{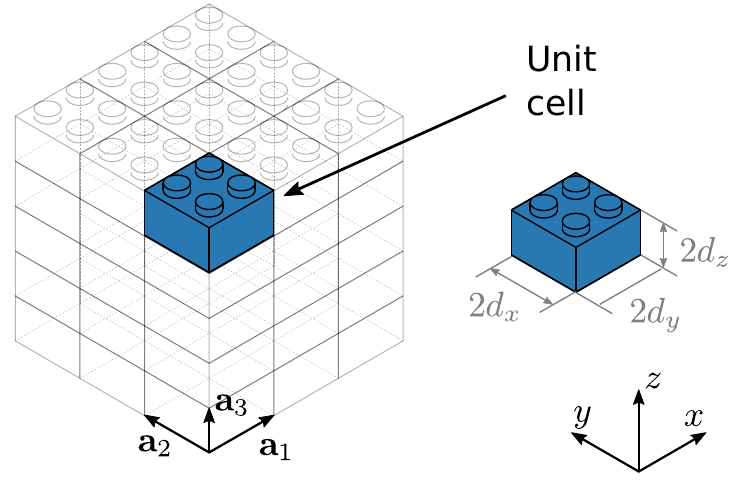}}
\caption{3D periodic material with different periodicities. Even though a material can be three dimensional, its periodicity could be (a) one, (b) two or (c) three dimensional. (d) Definition of the unit cell of a periodic material. The material is constructed applying translation operations to the unit cell following the lattice vector $\mathbf{a}$.}
\label{fig:lego}
\end{figure}

A periodic material is defined as the repetition of a given motif in one, two 
or three space dimensions. This motif refers to heterogeneities at 
micro-structural level and it may contain several materials and geometric 
features. \Cref{fig:lego}(a)-(c) show a three dimensional material with 
periodicity in one, two and three dimensions. Such periodic materials are 
completely described by a lattice and an elementary unit, termed 
\emph{elementary or unit cell}. The lattice is defined by a set of \emph{base 
vectors} (\cref{fig:lego}(d)), which allow construction of the whole material 
through successive applications of translation operations of the unit cell 
\citep{brillouin2003wave}. The occurrence of bandgaps in periodic materials is 
controlled by two fundamental mechanisms. Brag scattering appearing when the 
wavelength $\lambda$ of the propagating field assume values close to the 
characteristic size of the material miscroscructure and by local resonance 
induced by the combination of materials with strong impedance contrasts 
\citep{hussein2014}. The intrinsic periodicity of the material facilitates the 
characterization in terms of its dispersion relationships or band structure 
through Bloch's theorem as stated in \cite{brillouin2003wave} and discussed 
next.

\subsection{Bloch Theorem}
Let us consider a generalized wave equation in the frequency domain

\begin{equation}
  \mathcal{L} \mathbf{u}(\mathbf{x}) = -\rho \omega^2 \mathbf{u}(\mathbf{x})\, ,
  \label{eq:reduced}
\end{equation}

valid for a given field $\mathbf{u}$ at a spatial point $\mathbf{x}$ and where $\mathcal{L}$ is a positive definite differential operator 
\citep{book:reddy_functional,book:kreyszig_functional,johnson2007waves}, while $\rho$ is the mass density and $\omega$ the corresponding angular frequency. Bloch's theorem from solid state physics \citep{brillouin2003wave} establishes that the solution to \cref{eq:reduced} is given by
\begin{equation}
  \mathbf{u}(\mathbf{x}) = \mathbf{w}(\mathbf{x}) e^{i\mathbf{k}\cdot \mathbf{x}}\, ,
  \label{eq:bloch}
\end{equation}
where $\mathbf{w}(\mathbf{x})$ is a Bloch function carrying with it the same periodicity of the material. As a result the solution is the product of a periodic function, with the periodicity of the lattice, and a plane wave of wave vector $\textbf{k}$, which is also periodic. In consequence, field variables $\mathbf{\Phi}$ at opposite sides of the unit cell and separated by a vector $\mathbf{a}$ are related through

\[\mathbf{\Phi}(\mathbf{x} + \mathbf{a}) = \mathbf{\Phi}(\mathbf{x})e^{i\mathbf{k}\cdot\mathbf{a}}\, .\]

In this case, $\mathbf{\Phi}$ refers to the principal variable  involved in the physical problem, or to any of its spatial derivatives.

In the particular case in which $\mathcal{L}$ is of order 2, the generalized boundary value problem takes the form:

\begin{subequations}
\begin{align}
&\mathcal{L} \mathbf{u}(\mathbf{x}) = -\rho \omega^2 \mathbf{u}(\mathbf{x})\, ,\\
&\mathbf{u}(\mathbf{x} + \mathbf{a}) = \mathbf{u}(\mathbf{x})e^{i\mathbf{k}\cdot\mathbf{a}}\, ,\\
&\nabla \mathbf{u}(\mathbf{x} + \mathbf{a}) \cdot \hat{\mathbf{n}} = \nabla \mathbf{u}(\mathbf{x}) \cdot \hat{\mathbf{n}}\,  e^{i\mathbf{k}\cdot\mathbf{a}}\, ,
\end{align}
\label{eq:bvp_bloch}
\end{subequations}

where $\mathbf{u}(\mathbf{x} + \mathbf{a})$ and $\mathbf{u}(\mathbf{x})$ give the field at $\mathbf{x} + \mathbf{a}$ and $\mathbf{x}$ respectively while  $\mathbf{a} = \mathbf{a}_1 n_1 + \mathbf{a}_2 n_2 + \mathbf{a}_3 n_3$ is the lattice translation vector shown in \cref{fig:lego}(d). The term $e^{i\textbf{k}\cdot\textbf{a}}$ represents a phase shift between opposite sides of the unit cell. This relationship between opposite sides of the fundamental cell stated in the theorem through the boundary terms permits the characterization of the fundamental properties of the material with the analysis of a single cell.  The section that follows describes the theorem within the particular context of a finite element formulation following standard Galerkin ideas.

\subsection{Finite element formulation}

\begin{figure}[H]
\centering
\includegraphics[width=15cm]{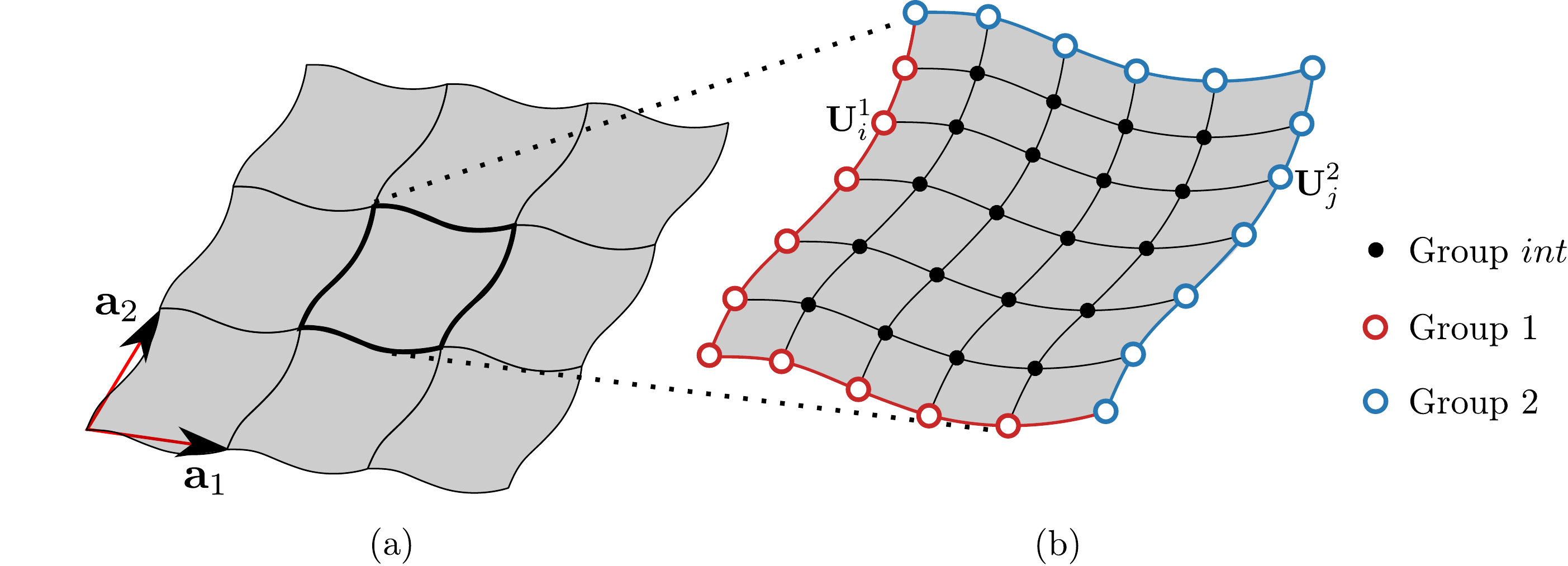}
\caption{(a) Schematic 2D periodic material and its lattice translation vectors $\textbf{a}_1$ and $\textbf{a}_2$. (b) FE mesh from the unit cell and division in groups of nodes. Nodes in group 2 (blue) are equivalent to nodes in group 1 (red) in an adjacent unit cell. A pair of nodes are equivalent if their coordinates can be expressed as $\textbf{x}_j^2 =  \textbf{x}_i^1 + m\textbf{a}_1 + n\textbf{a}_2$ for integers $m$ and $n$. }
\label{fig:gmesh}
\end{figure}

\Cref{fig:gmesh}a shows an schematic representation of the unit cell in a periodic material. The corresponding finite element discretization of the BVP stated in \cref{eq:bvp_bloch} as shown in the mesh in \cref{fig:gmesh}b takes the form

\begin{equation}
\left[K - \omega^2M\right] \{ \textbf{U} \} = \textbf{0}
\label{eq:eigG}
\end{equation}

where in the context of elastodynamics, $K$ and $M$ would correspond to global stiffness and mass matrices respectively. Note that \cref{eq:eigG} constitutes an eigenvalue problem in the eigenvalues $\omega$. The schematic mesh contains also two different sets of nodal points corresponding to (i) interior nodes, labeled as $int$; and (ii) exterior nodes, labeled as 1 and 2. Nodes from groups 1 and 2 are represented in red and blue colors respectively. Imposition of Bloch periodic boundary conditions to the nodal sets 1 and 2 is dictated by the boundary conditions specified in \cref{eq:bvp_bloch}b, \citep{guarin2014evaluation}.  In general, two equivalent nodes $i,j$ in different adjacent cells can be related with the BBCs by

\begin{equation}
\textbf{U}_j^2 = \textbf{U}_i^1 \times \textbf{PS}_{i \rightarrow j}
\label{eq:bbc}
\end{equation}

where $\textbf{U}_i^1$ is the field of node $i$ belonging to group 1, $\textbf{U}_j^2$ is the field of node $j$ (equivalent to $i$ in a different periodic cell) belonging to group 2, and $\textbf{PS}_{i \rightarrow j}$ is the phase shift between the nodes $i$ and $j$, with $\textbf{PS}_{i \rightarrow j} = e^{i\mathbf{k}\cdot\mathbf{a}}$. The imposition of the BBCs to the complete system and the subsequent removal of redundant equations belonging to nodes in group 2, yields the reduced version of \cref{eq:eigG} in terms of reduced matrices $K_R$ and $M_R$ as:

\begin{equation}
\left[K_R - \omega^2M_R\right] \{ \textbf{U}_R \} = \textbf{0} \quad .
\label{eq:eig}
\end{equation}

\begin{figure}
\centering
\includegraphics[width=11 cm]{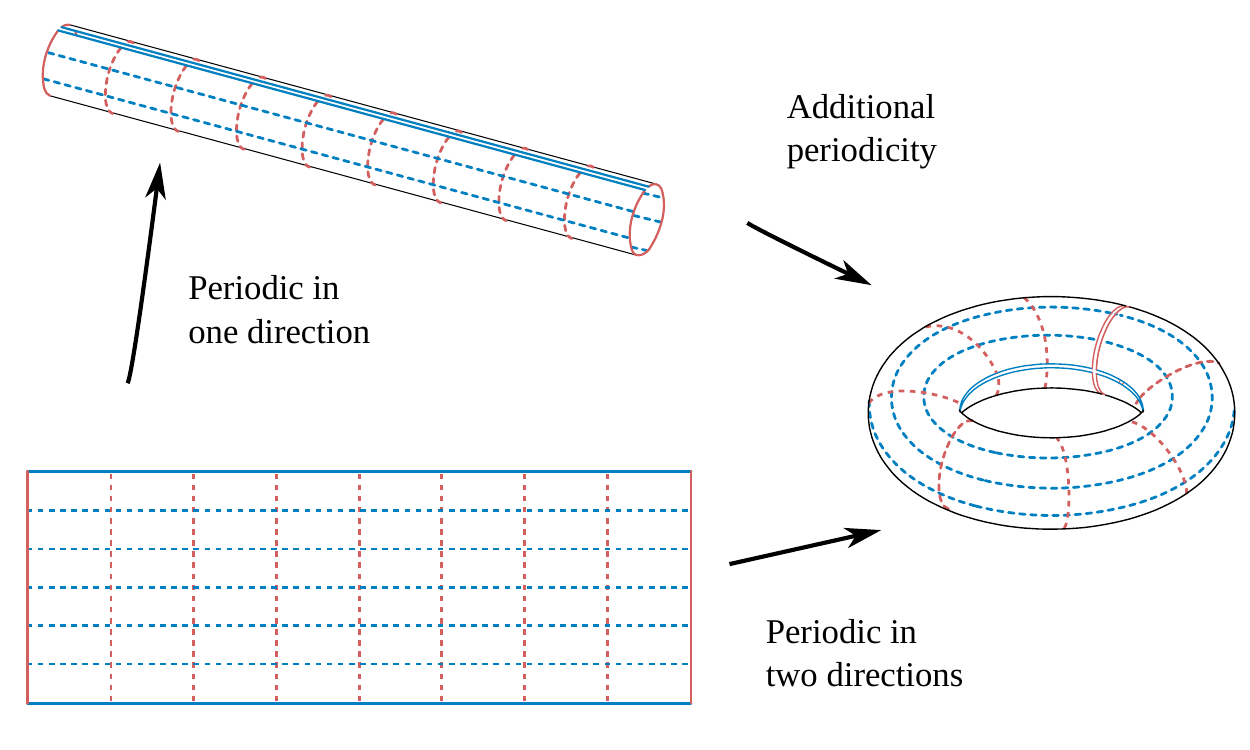}
\caption{Topology for a rectangular region with periodicity in two directions. When periodicity in one direction is applied the mesh can be thought as laying on the surface of a cylinder. In the case of periodicity in two direction the mesh can be thought as laying on the surface of a torus.}
\label{fig:topology}
\end{figure}

The periodicity condition is illustrated in \cref{fig:topology} for the 
periodic material displayed in the rectangular domain. Periodicity along the 
vertical direction is equivalent to folding the rectangular sheet onto itself 
forming the cylinder at the top. Similarly, considering also periodic 
variations in the horizontal direction converts the cylinder into a torus as 
shown in the right. Note that $K_R$ and $M_R$ in \cref{eq:eig} depend upon the 
wave vector $\textbf{k}$. Accordingly, for a given $\textbf{k}$ \cref{eq:eig} 
gives a particular instance of the eigenvalue problem. The band structure of 
the material is thus built after progressively covering the first Brillouin 
zone \citep{brillouin2003wave} in the wave-number domain representation of the 
unit cell. Each solution to the generalized eigenvalue problem given in 
\cref{eq:eig} represents a free wave propagating at frequency $\omega$, 
moreover each corresponding solution gives all frequencies $\omega$ at which 
propagation of the specific free wave is possible and the dispersion relation 
$\omega$ vs $\textbf{k}$ is then constructed.

\subsection{Imposition BBCs in global matrices}

\begin{figure}[H]
\centering
\subfloat[Meshed unit cell]{\includegraphics[width=0.25\textwidth]{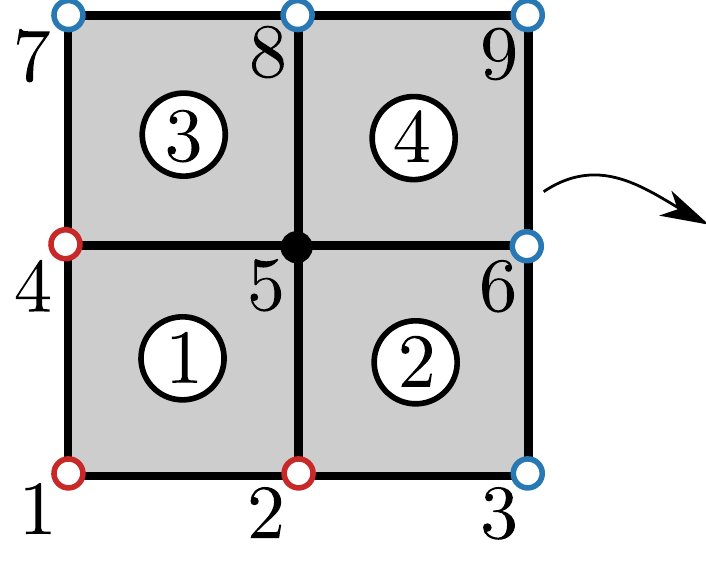}
\label{fig:mesh}}
\subfloat[Global matrix scheme]{\includegraphics[width=0.35\textwidth]{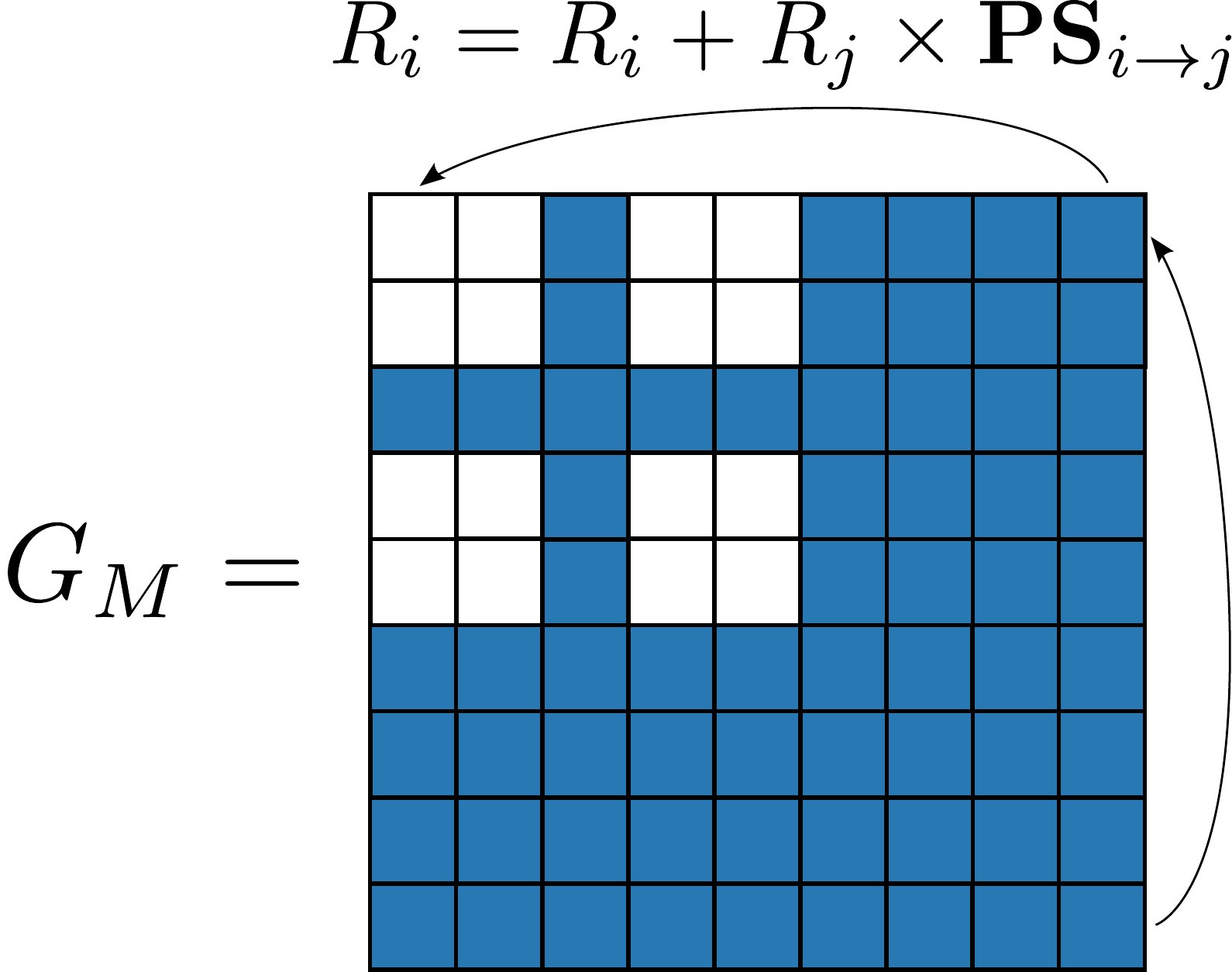}\quad
\label{fig:gm}}
\subfloat[Reduced matrix scheme]{\includegraphics[width=0.35\textwidth]{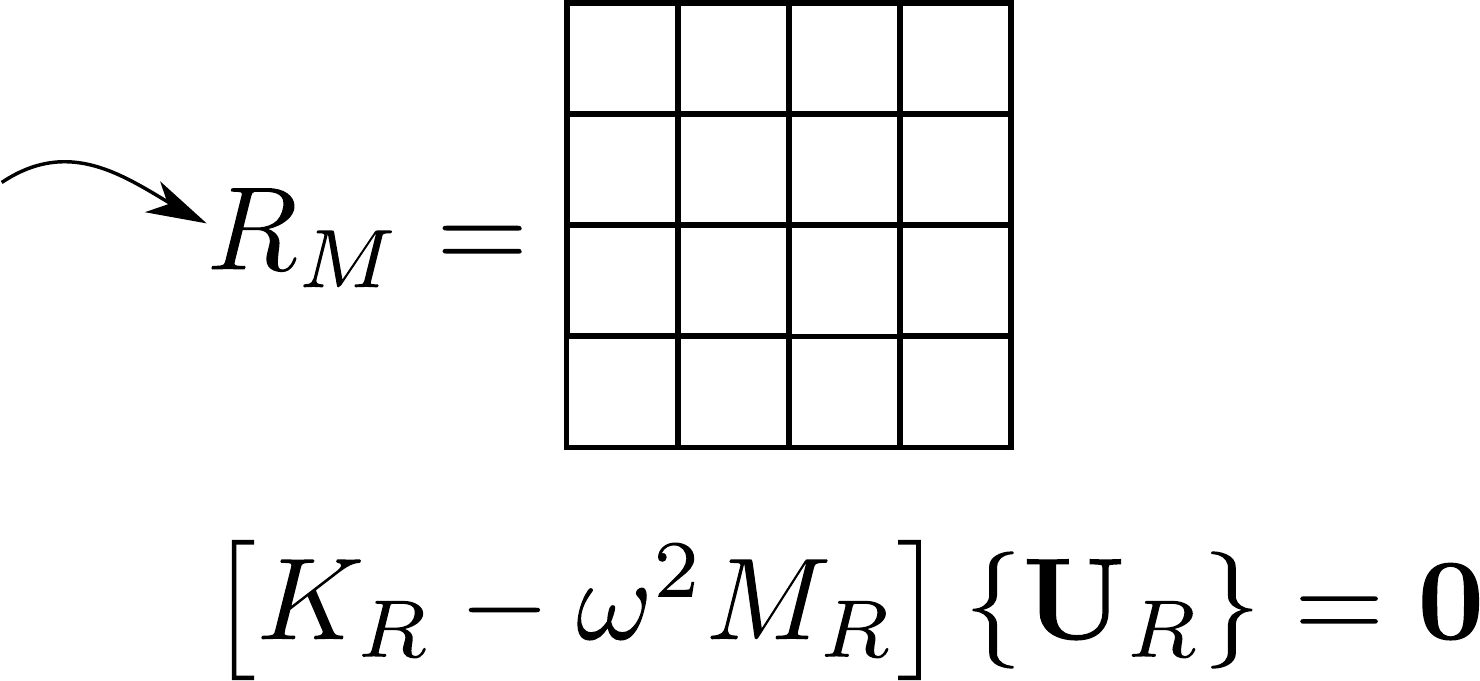}
\label{fig:rm}}\\
\caption{Schematic of the reduced eigenvalue problem after applying BBCs to the global matrices. The squares in the global matrix represent sub--matrices with equations from nodes in the mesh. After application of BBCs, blue squares collapse into the white squares. Namely, equations from nodes in group 2 collapse into equations from nodes in group 1. Finally, the redundant (blue) equations are removed from the matrices resulting in a reduced set of matrices.}
\label{fig:exam}
\end{figure}

\Cref{fig:mesh} shows a 4-bilinear-elements mesh corresponding to the discretization of a fundamental unit cell in a periodic material. The discrete equations from the general mesh are assembled into global stiffness and mass matrices $K$ and $M$ respectively as shown in the graphic description from \cref{fig:gm} and where each square section represents the sub-matrix associated to the set of degrees of freedom (DOF) at a specific node. In this case, and just for illustration purposes, we assume 1-DOF per node as in a scalar problem so each element contributes to the global system with $4 \times 4$ global matrices $G_M$.

Now, imposition of the required BBCs implies collapsing into a single equation 
those of nodes related as in \cref{eq:bbc}. For instance, nodes 9 and 1 at 
opposite corners of the unit cell satisfy:

\begin{equation}
R_1 = R_1 + R_9 \times \textbf{PS}_{1 \rightarrow 9},
\end{equation}

\noindent
with $R_n$ denoting a row (or a column) of sub-matrices in the general system 
$G_M$ associated to node $n$. This means that equations from node 9 will 
collapse into those from node 1 after imposition of BBCs. Final application of 
BBCs to all relevant nodes through a process of removing redundant equations 
from $K$ and $M$ leads to reduced matrices $K_R$ and $M_R$ forming the 
generalized eigenvalue problem given in \cref{eq:eig}. With reference to the 
schematic example of \cref{fig:exam} note that equations from nodes $3,6,7,8,9$ 
(blue squares in \cref{fig:gm}) were properly collapsed into those of nodes 
$1,2,4$ (white squares in \cref{fig:gm}) producing the reduced matrices in 
\cref{fig:rm}. This row-and-column operation scheme is widely used in Bloch 
analysis but it is difficult if not possible to apply in finite element codes 
that restrain access to the global coefficient matrices. Some available codes 
allow the manipulation to the global system as required in Bloch analysis 
through the imposition of Multi-Point constraints 
\citep{hibbett1998abaqus,mazzoni2006opensees}. In these cases however, the 
analysis is restricted to the physical context available in the code.

\section{Simple procedure to imposition of BBCs with user element subroutines }

\label{sec:uel}
We now introduce our alternative approach to implement BBCs as required in the analysis of periodic media. The proposed approach does not require access to the global arrays but instead it conducts all modifications at the element level. As such, the method is suitable for codes allowing the implementation of user element subroutines. In these codes, the main program acts as a solver of a system of equations where the user provides the contribution from each element to the global system. Here we take advantage of these features and instead of conducting the row-column operations leading to $K_R$ and $M_R$ onto the global arrays we compute these matrices directly at the element level just by varying the standard assembly procedure available in FE codes.

\subsection{Imposition BBCs in elemental matrices }
To clarify let us denote the set of connectivity nodes for a typical element $m$ as $C_m$. Generally, $C_m$ is used to carry out two tasks: (1) to define the coordinates of the nodes in element $m$ which are required in the computation of elemental matrices; (2) to assemble these elemental matrices into the global arrays at positions defined by the equations identifiers associated to these nodal points. In a classic FE problem those two tasks are performed with the same nodal set $C_m$, however in the case of periodic materials carrying with it BBCs the assembly process proceeds differently. Here we define two kinds of connectivity operators $C_m$ corresponding to the classical coordinate-connectivity $C_m^c$ and an assemblage-connectivity set $C_m^a$. The first operator is used to compute the elemental matrices as in the standard approach while the second is used to conduct the assembly operation thus delivering global matrices where Bloch boundary conditions have already been applied depending on the definition of $C_m^a$ . In this approach $C_m^c$ and $C_m^a$ are defined only for those elements containing nodes belonging to group 2 as described in \cref{fig:gmesh} while in the remaining elements $C_m^c$ and $C_m^a$ are identical.

\begin{figure}[H]
\centering
\includegraphics[width=12cm]{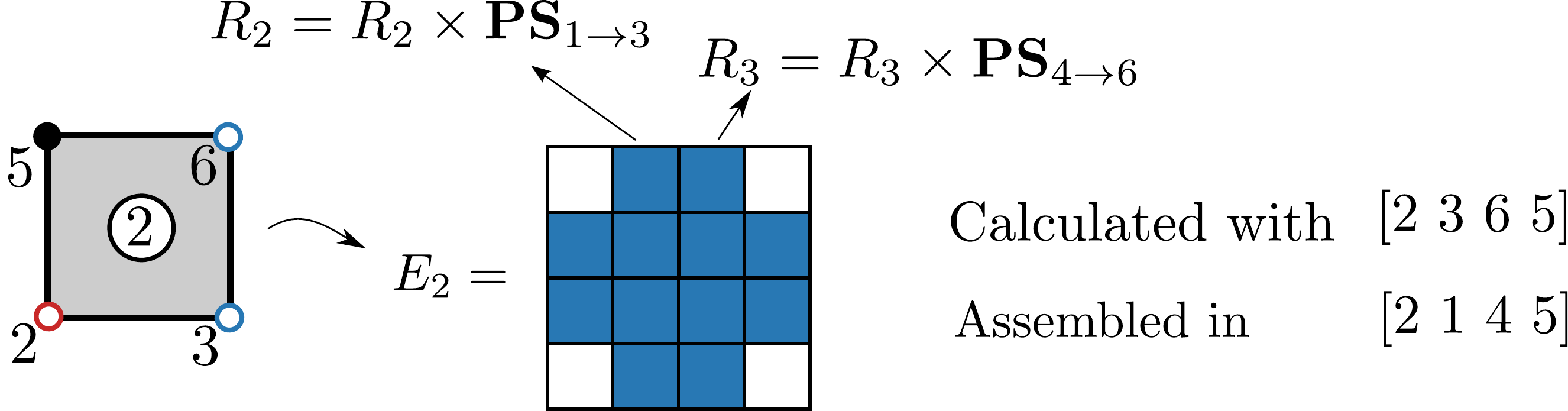}
\caption{Application of BBC to local matrices from element 2. Equations from nodes 3 and 6 are represented as blue sub-matrices. They include their proper phase shift and will be assembled in the positions dictated by $C_2^a = [$2 1 4 5$]$.}
\label{fig:uel}
\end{figure}

Consider element 2 in the mesh shown in \cref{fig:mesh}. Due to the periodicity for this specific cell, nodal pairs 6 and 4 and 3 and 1 are related by BBCs in such a way that in the global matrices, equations from node 6 will collapse into those from node 4, while equations from node 3 will collapse into equations from node 1. As a result we define connectivity operators $C_2^c=[2$ $3$ $6$ $5]$ and $C_2^a=[2$ $1$ $4$ $5]$. This process must also incorporate the phase shifts corresponding to the impositions $\textbf{PS}_{1 \rightarrow 3}$ and $\textbf{PS}_{4 \rightarrow 6}$ to the equations of nodes 3 and 6 respectively in the local matrix. At the end of the procedure, equations from nodes in element 2 would have the proper phase shift and would be assembled in the proper global positions as pointed out by $C_2^a$. This process is explained graphically in \cref{fig:uel}.

The next step in the assembly of the final global arrays is the elimination of 
redundant equations (in this case those in nodes 3, 6, 7, 8, 9) to obtain 
reduced arrays $K_R$ and $M_R$. Here this is achieved by restraining the 
degrees of freedom associated to these redundant equations. That process 
guarantees that $K_R$ and $M_R$ are assembled considering the required BBCs and 
also that they are ready for solution of the generalized eigenvalue problem. 
Since this strategy to apply BBCs to a discretized unit cell proceeds at the 
element level and it is independent of the problem at hand it can be 
straightforwardly used for one-, two- or three-dimensional periodic materials, 
using any kinematic model or interpolation scheme. Furthermore, it can be 
easily added to existing user element subroutines by making just subtle changes.

\subsection{Reformulation in real algebra}

An additional difficulty that appears often when conducting Bloch analysis with commercial finite element codes is the fact that the resulting generalized eigenvalue problem given by \cref{eq:eig} is a complex-valued system. This is a consequence of the phase shifts of the general form $e^{i\mathbf{k}\cdot\mathbf{a}}$ explicitly appearing in BBCs. Most codes have powerful and efficient real-algebra-based built-in eigensolvers. To take advantage of these numerical features we have followed \cite{aaberg1997usage}, who split \cref{eq:eig} into two real problems as elaborated next.

\begin{figure}[H]
\centering
\includegraphics[width=13cm]{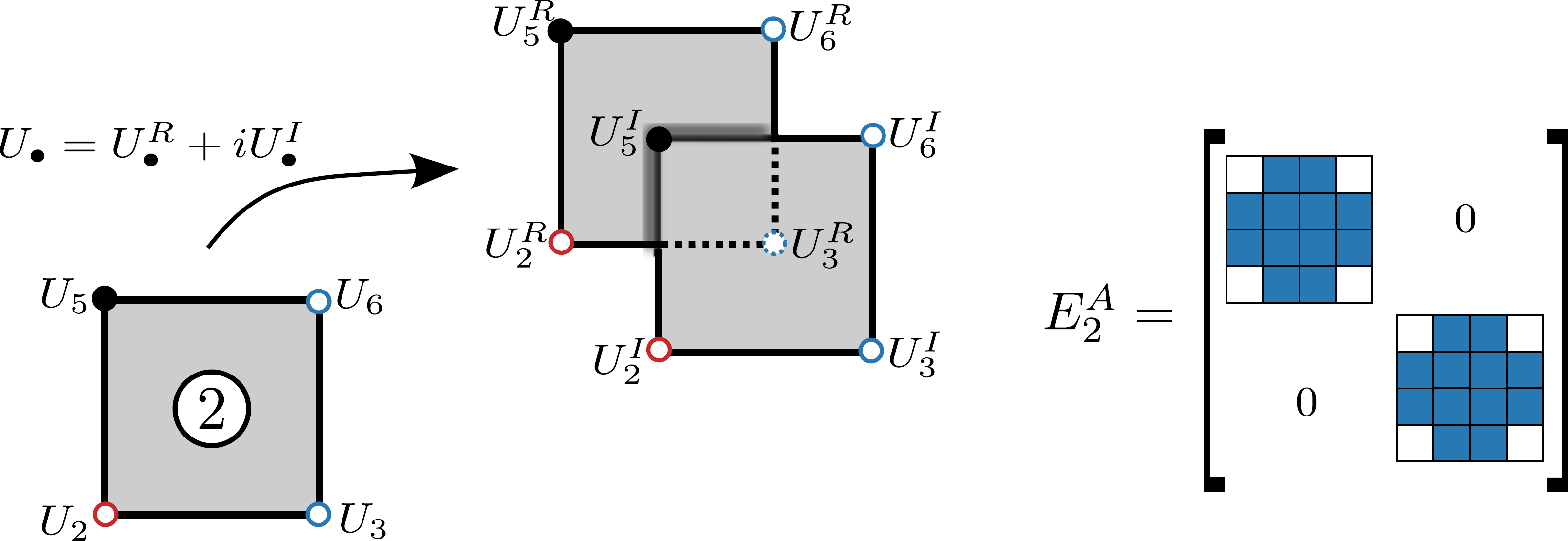}
\caption{Application of BBCs to element 2 following the real algebra splitting strategy proposed by \cite{aaberg1997usage}. The equations from the element are split into real and imaginary parts, producing an augmented elemental matrix. BBCs are the applied to both parts of the new matrix.}
\label{fig:uel_r}
\end{figure}

Consider once again the 4-noded quadrilateral element showed in \cref{fig:uel} and repeated with additional information in \cref{fig:uel_r}. Recall that nodal pairs 6 and 4 are related by $U_6 = U_4\textbf{PS}_{4 \rightarrow 6}$, with $\textbf{PS}_{4 \rightarrow 6} = e^{i\mathbf{k}\cdot \mathbf{a}}$ and where $U_6$ and $U_4$ are the set of equations from nodes 6 and 4 respectively. These are complex valued terms which can be spitted like $U_\bullet = U_\bullet^R + i U_\bullet^I$ with $U_\bullet^R$ and $U_\bullet^I$ being respectively the real and imaginary components of $U$ while $i$ is the imaginary unit. The original mesh can be interpreted now as a duplicated mesh where each part handles the real and imaginary component of the equations. This is explained in  \ref{fig:uel_r} where we show the full-matrix $E_2^A$ composed of the double consideration of the matrix $E_2$ like:

\begin{equation*}
E_2^A
=
\begin{bmatrix}
    [E_2] & 0 \\
    0 & [E_2]
\end{bmatrix}
\end{equation*}

Consideration of all nodes in the element satisfying Bloch periodic boundary 
conditions, leads to a relation between real and imaginary meshes as stated by 
\cite{aaberg1997usage} and given by:

\begin{subequations}
\begin{eqnarray}
U_6^R = U_4^R \cos \left(\mathbf{k}\cdot \mathbf{a}\right) + U_4^I \sin \left(\mathbf{k}\cdot \mathbf{a}\right) \\
U_6^I = U_4^I \cos \left(\mathbf{k}\cdot \mathbf{a}\right) - U_4^R \sin \left(\mathbf{k}\cdot \mathbf{a}\right)  \\
U_3^R = U_1^R \cos \left(\mathbf{k}\cdot \mathbf{a}\right) + U_1^I \sin \left(\mathbf{k}\cdot \mathbf{a}\right) \\
U_3^I = U_1^I \cos \left(\mathbf{k}\cdot \mathbf{a}\right) - U_1^R \sin \left(\mathbf{k}\cdot \mathbf{a}\right)
\end{eqnarray}
\label{eq:bbcr}
\end{subequations}

\begin{figure}[H]
\centering
\includegraphics[width=14cm]{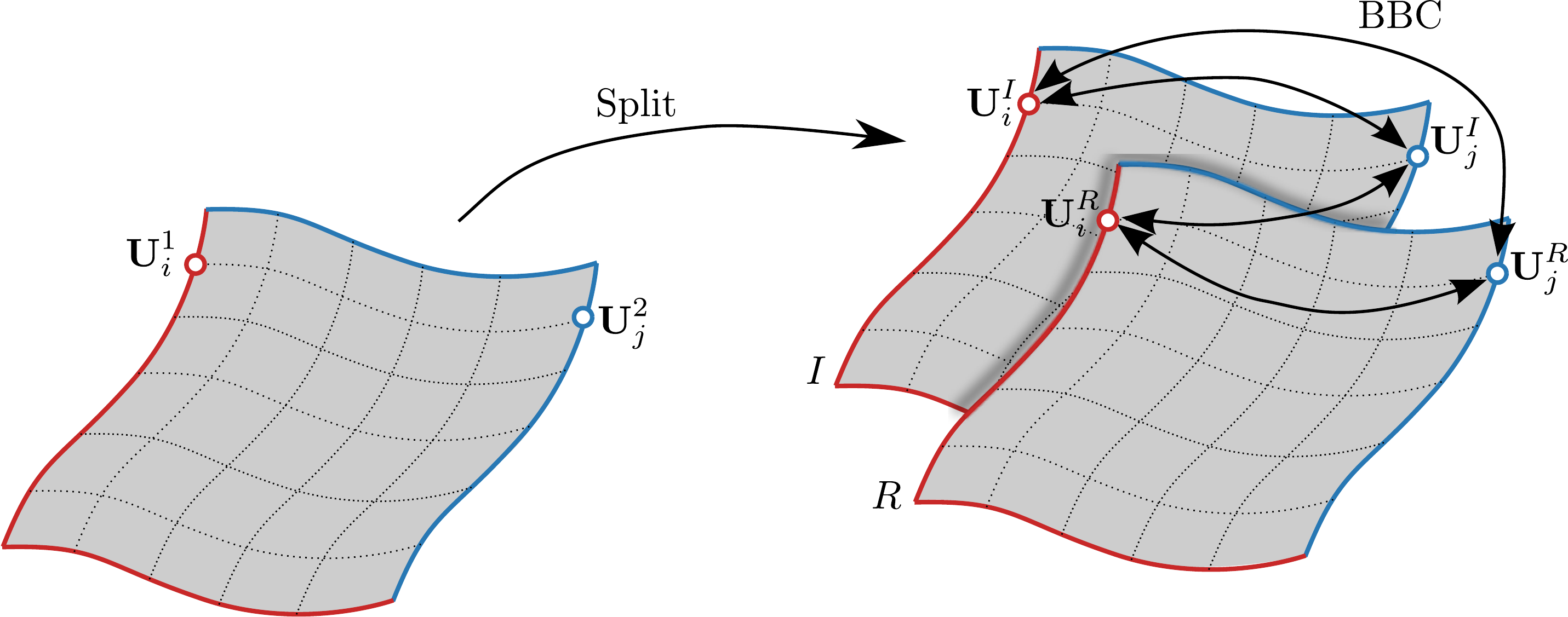}
\caption{Schematic depiction of the splitting process. The mesh and the BBCs are split into real part and imaginary part. Real and imaginary parts from BBCs are combined and applied to each part of the mesh following procedure similar to the one in \cref{eq:bbcr} for every pair of related nodes. The new mesh can be thought as two identical meshes attached by the borders.}
\label{fig:uel_rc}
\end{figure}

These sets of DOF modified by BBCs are represented in the schematic matrix representation of \cref{fig:uel_r} as sub-matrices in blue rows and columns. Those specific sub-matrices are to be assembled in the positions given by $C_2^a=[2$ $1$ $4$ $5]$. Note that although BBCs, like those in  \cref{eq:bbcr} are just applied to local matrices from elements containing at least one node from group 2 the dual-mesh representation has to be applied for all elements in the mesh. According to the color scheme in \cref{fig:uel_r}, local matrices from elements containing any node from group 2,  have blue equations; while the rest of elements have white equations. The final result is therefore  a duplicated mesh where real and imaginary parts from BBCs are applied to both real and imaginary portions of the mesh. \Cref{fig:uel_rc} shows an schematic illustration of the splitting process over a general mesh and the application of BBCs to the global terms of the mesh. The final assembly can be thought as corresponding to two identical superimposed meshes, where the blue border collapses into the red border. Finally the positions from the blue border are removed from the global arrays as indicated previously.

\section{Verification}

To test the generality of our implementation we computed the dispersion 
relations in the context of elastodynamics, for three class of media with 
different kinematic assumptions, as described below. However it must be kept in 
mind that the proposed methodology to impose BBCs is independent of the 
physical context of the problem at hand and it can be applied to scalar 
problems (e.g., linear acoustics), to quantum mechanics problems, or to vector 
problems (e.g.,  electrodynamics or elastodynamics). For waves propagating in 
unbounded domains we can consider the problem as two-dimensional where the wave 
polarization can be in or out-of-plane \citep{book:achenbach, book:auld1973}. 
In this study we consider waves propagating in the $xy$ plane for the following 
cases:

\paragraph{Out-of-plane waves in elastic media:} The simplest case correspond to out-of-plane horizontally polarized shear or $SH$ waves. These are governed by displacement equations of motion involving only the displacement in the $z$-direction $u_z$ with reduced wave equation of the form:
\begin{equation*}
  \frac{\mu}{\rho} \left[\pdv[2]{u_z}{x} + \pdv[2]{u_z}{y}\right] = -\omega^2 u_z\, ,
  \label{eq:wave_sh}
\end{equation*}
where $\mu$ and $\rho$ are the second Lamé parameter (or shear modulus) and the volumetric mass density defining the phase speed $c_T^2 = \mu/\rho$.

\paragraph{In-plane waves in elastic media:} The vectorial displacement field in elastic media is completed considering in-plane polarization in terms of two degrees of freedom, corresponding to the horizontal and vertical components $u_x$ and $u_y$ at each material point. The wave motion in this case is described by the following specific form of the reduced wave equation:
\begin{align*}
&\frac{\lambda + 2\mu}{\rho}\left[\pdv[2]{u_x}{x} + \pdv{u_y}{y}{x}\right] - \frac{\mu}{\rho}\left[\pdv{u_y}{y}{x} - \pdv[2]{u_x}{y}\right] = -\omega^2 u_x\, ,\\
&\frac{\lambda + 2\mu}{\rho}\left[\pdv{u_x}{y}{x} + \pdv[2]{u_y}{y}\right] - \frac{\mu}{\rho}\left[\pdv{u_x}{y}{x} - \pdv[2]{u_y}{x}\right]= -\omega^2 u_y\, ,
\end{align*}
where $\lambda$ is the first Lamé parameter, $\mu$ is the second Lamé parameter or shear modulus, and $\rho$ is the volumetric mass density. In this case we have two wave modes with phase speeds $c_L^2 = (\lambda + 2\mu)/\rho$ and $c_T^2 = \mu/\rho$.

\paragraph{In-plane waves in micropolar media:} Micropolar media, inspired by 
contributions from \cite{cosserat1909theorie}, correspond to a continuum model 
where each material point is endowed with six degrees of freedom in the form of 
three displacements --- as in classical elasticity --- and three rotations. In 
this resulting micropolar elasticity theory the transmission of loads through 
surface elements is now described by force and couple stress vectors. As a 
consequence of these additional kinematic interactions, waves in (homogeneous, 
isotropic) micropolar continua are inherently dispersive even in the absence of 
explicit micro-structural details. Under two-dimensional idealizations 
micropolar media is defined in terms of two displacements and one rotation 
leading to the following equations representative of the wave motion for such a 
class of media:
\begin{align*}
&\frac{\lambda + 2\mu}{\rho}\left[\pdv[2]{u_x}{x} + \pdv{u_y}{y}{x}\right] - \frac{\mu + \mu_c}{\rho}\left[\pdv{u_y}{y}{x} - \pdv[2]{u_x}{y}\right] + \frac{2\mu_c}{\rho}\pdv{\phi_z}{y} = -\omega^2 u_x\, ,\\
&\frac{\lambda + 2\mu}{\rho}\left[\pdv{u_x}{y}{x} + \pdv[2]{u_y}{y}\right] - \frac{\mu + \mu_c}{\rho}\left[\pdv{u_x}{y}{x} - \pdv[2]{u_y}{x}\right] - \frac{2\mu_c}{\rho}\pdv{\phi_z}{y} = -\omega^2 u_y\, ,\\
&\frac{2\mu_c}{J}\left[\pdv{u_y}{x} - \pdv{u_x}{y}\right] + \frac{\xi}{J}\left[\pdv[2]{\phi_z}{x} + \pdv[2]{\phi_z}{y}\right] - \frac{4\mu_c}{J}\phi_z = -\omega^2 \phi_z\, ,
\end{align*}
and where in addition to the classical mechanics first and second Lamé 
parameters $\lambda$ and $\mu$ there are also new moduli $\mu_c$ and $\xi$ with 
no classical counterparts Also in these equations $\rho$ and $J$ are the 
volumetric and rotational mass density. Under this non-classical kinematic 
model there are three wave modes with a volumetric wave with phase speed $c_L^2 
= (\lambda + 2\mu)/\rho$ and two dispersive modes with phase speeds that depend 
on the frequency \citep{book:nowacki1986}.

All of the above kinematic models are particular instances of \cref{eq:bvp_bloch}a.  In this study we take as base materials aluminum and brass with the following set of material properties:
\begin{table}[H]
\begin{center}
\begin{tabular}{ccccccc}
\cline{2-7}
& \multicolumn{6}{ c }{Material properties}  \\
& $\lambda$ (Pa) & $\mu$ (Pa) & $\rho$ (kg/m$^3$) & $\mu_c$ (Pa) & $\xi$ (N) & $J$ (kg/m) \\ \hline
\multicolumn{1}{ c  }{\textbf{Material 1}}  &
\multicolumn{1}{ c }{$5.12 \times 10^{10}$} & {$2.76 \times 10^{10}$} & 2770 & {$3.07 \times 10^{9}$} & {$7.66 \times 10^{9}$} & 306.5 \\ 
\multicolumn{1}{ c  }{\textbf{Material 2}} &
\multicolumn{1}{ c }{$6.45 \times 10^{10}$} & {$3.47 \times 10^{10}$} & 8270 & {$8.65 \times 10^{9}$} & {$1.73 \times 10^{9}$} &691.7 \\ \hline 
\end{tabular}
\caption{Mechanical properties for the materials used.}
\label{tab:prop}
\end{center}
\end{table}

\subsection{Test of generality about physics}

The first problem is a comparison between numerical and analytical results for a homogeneous material unit cell aimed at testing the capabilities of our strategy to describe new waves as we increase the complexity of the kinematic model. We computed the dispersion relations using the three different kinematic assumptions described above where in each case there are additional degrees of freedom added to the model.

\subsubsection{Analytical dispersion relations for a homogeneous material}

Dispersion relationships for the homogeneous material models can be written solely as functions of the magnitude of the wavenumber like
\[\omega \equiv \omega(k)\, .\]

However when these relationships are obtained from Bloch's theorem the dispersion relationships also provide  information from different Brillouin zones leading to relations of the form:

\begin{equation}
  \omega_{m, n} \equiv \omega(k_{m, n})\, ,
  \label{eq:disp_general}
\end{equation}
where the subscripts $m, n$ correspond to integer numbers making reference to waves coming from adjacent Brillouin zones. In the case of a square unit cell we have the following generalized definition of the wave number \citep{thesis:langlet}:
\begin{equation}
  k_{m,n} = \sqrt{\left(k_x + \frac{m\pi}{d}\right)^2 + \left(k_y + \frac{n\pi}{d}\right)^2}\, .
  \label{eq:wavenumber_adjacent}
\end{equation}

\paragraph{Classical elastodynamics:} 
In classical elastodynamics, described separately in terms of $SH$ out-of-plane and $P$, $SV$ in-plane waves, the dispersion relations for a homogeneous material cell take the analogous linear forms in terms of phase speeds $c_T/c_L$ as given by:

\begin{align*}
 &\omega^{SH}_{m, n} =  c_T k_{m, n}\, ,\\
  &\omega^P_{m, n} =  c_L k_{m, n}\, ,\\
  &\omega^{SV}_{m, n} =  c_T k_{m, n}\, ,
\end{align*}

\paragraph{Micropolar elastodynamics:} In this non-classical model there are three in-plane propagation modes. In addition to the P- and S-waves there is also a transverse rotational wave (TR). Furthermore, in this case S-waves are dispersive as can be seen in the following dispersion relations:
\begin{align*}
  &\omega^P_{m, n} =  c_L k_{m, n}\, ,\\
  &\omega^S_{m, n} =  \sqrt{\frac{A}{2} - \frac{1}{2}\sqrt{A^2 - 4B}}\, ,\\
  &\omega^{TR}_{m,n} = \sqrt{\frac{A}{2} + \frac{1}{2}\sqrt{A^2 - 4B}}\, ,
\end{align*}
with
\begin{align*}
  &A = 2Q^2 + (c_2^2 + c_4^2) k_{m, n}^2\, ,\\
  &B = 2Q^2 c_2^2 k_{m, n}^2 - K^2 Q^2 k_{m, n}^2 + c_2^2 c_4^2 k_{m, n}^4\, ,
\end{align*}
and
\begin{equation*}
\begin{split}
c_2^2 = \frac{\mu + \mu_c}{\rho},\quad &c_4^2 =\frac{\xi}{J},\\
K^2= \frac{2\alpha}{\rho},\quad &Q^2 =\frac{2\alpha}{J} \, .
\end{split}
\end{equation*}

\subsubsection{Numerical results}

To test the proposed strategy we implemented the elemental-based approach for the kinematic models described previously into the finite element code FEAP, \citep{taylor2011feap}. In the supplementary material of this work we added a fully-coded version of the algorithm as a FEAPpv binary file together with a fully defined example problem to test it. The files also include subroutines to read the required $C_m^c$ and $C_m^a$ operators and peripheral codes to cover the cell representation in the reciprocal wave number domain. All the dispersion graphs use the dimensionless frequency
\begin{equation}
  \Omega = \frac{2 \omega d}{c_T}\, ,
  \label{eq:dimensionless_freq}
\end{equation}
for the vertical axis, where $2d$ is the dimension of the unit cell and $c_T^2 = \mu/\rho$ is the speed of the shear wave.

\begin{figure}
\centering
\begin{tabular}{m{0.27\textwidth}m{0.27\textwidth}m{0.27\textwidth}m{0.1\textwidth}}
\subfloat[Out-of-plane]{\includegraphics[width=0.3\textwidth]{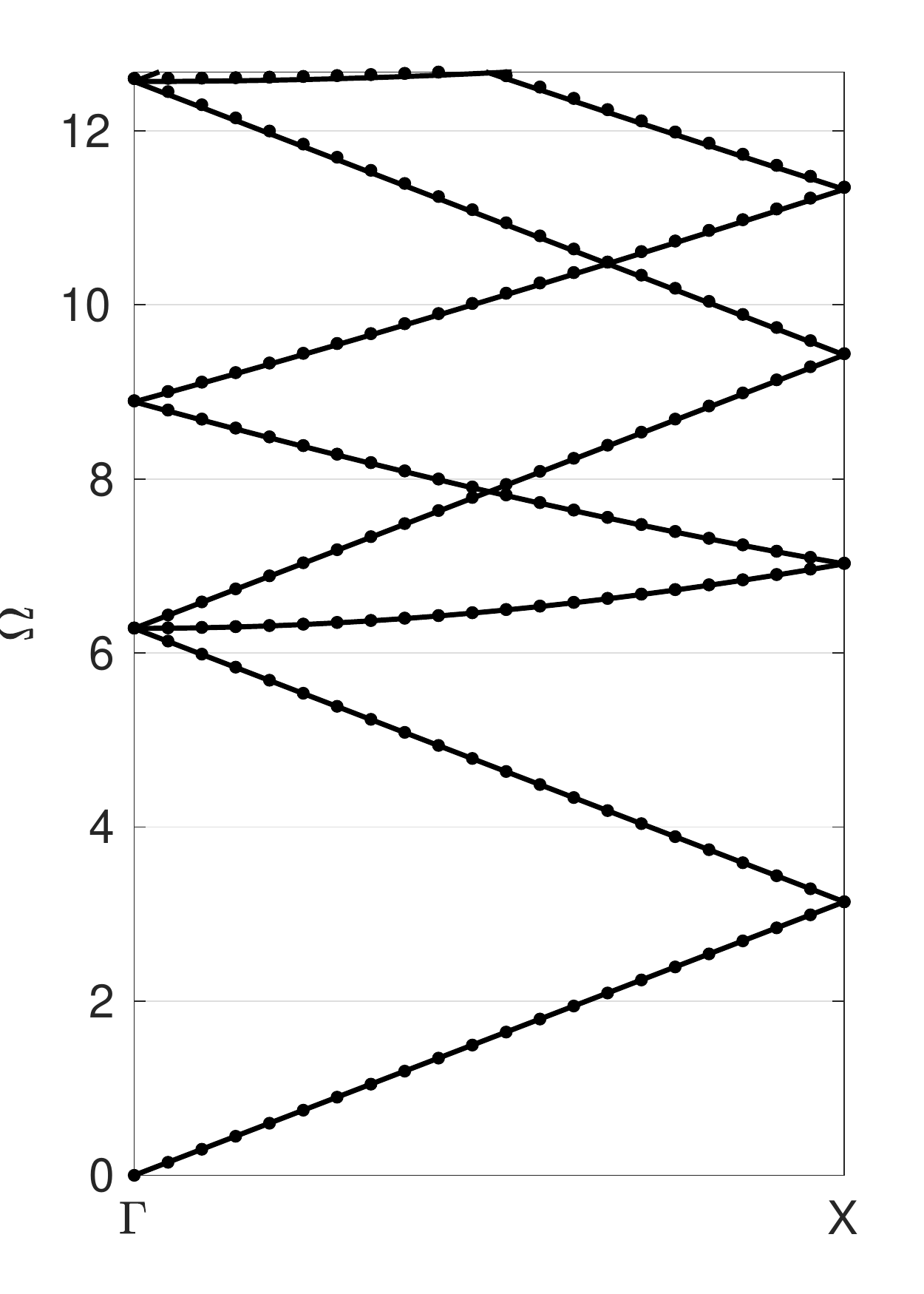}
\label{fig:homosh}} &
\subfloat[In-plane]{\includegraphics[width=0.3\textwidth]{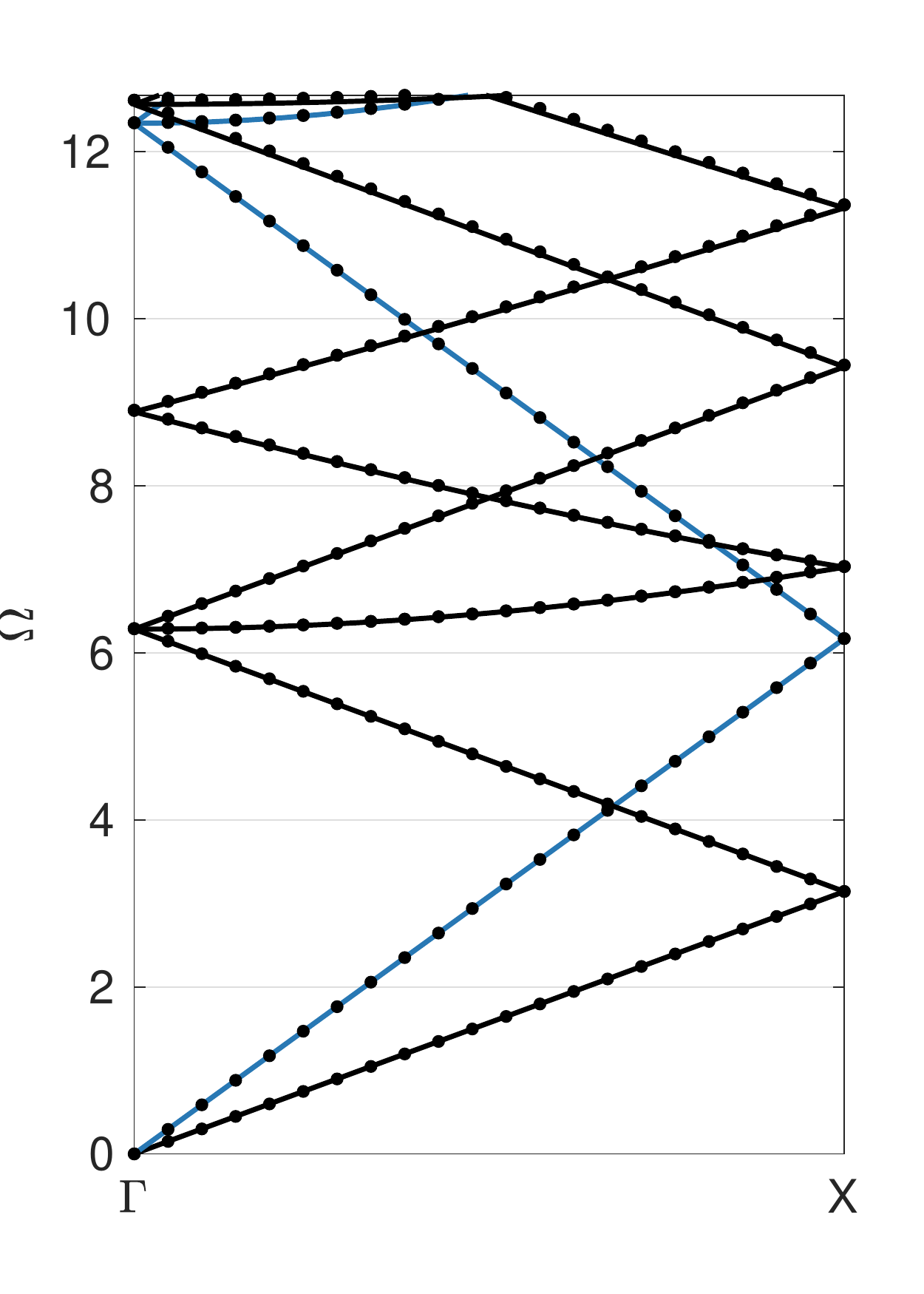}
\label{fig:homopsv}} &
\subfloat[Micropolar]{\includegraphics[width=0.3\textwidth]{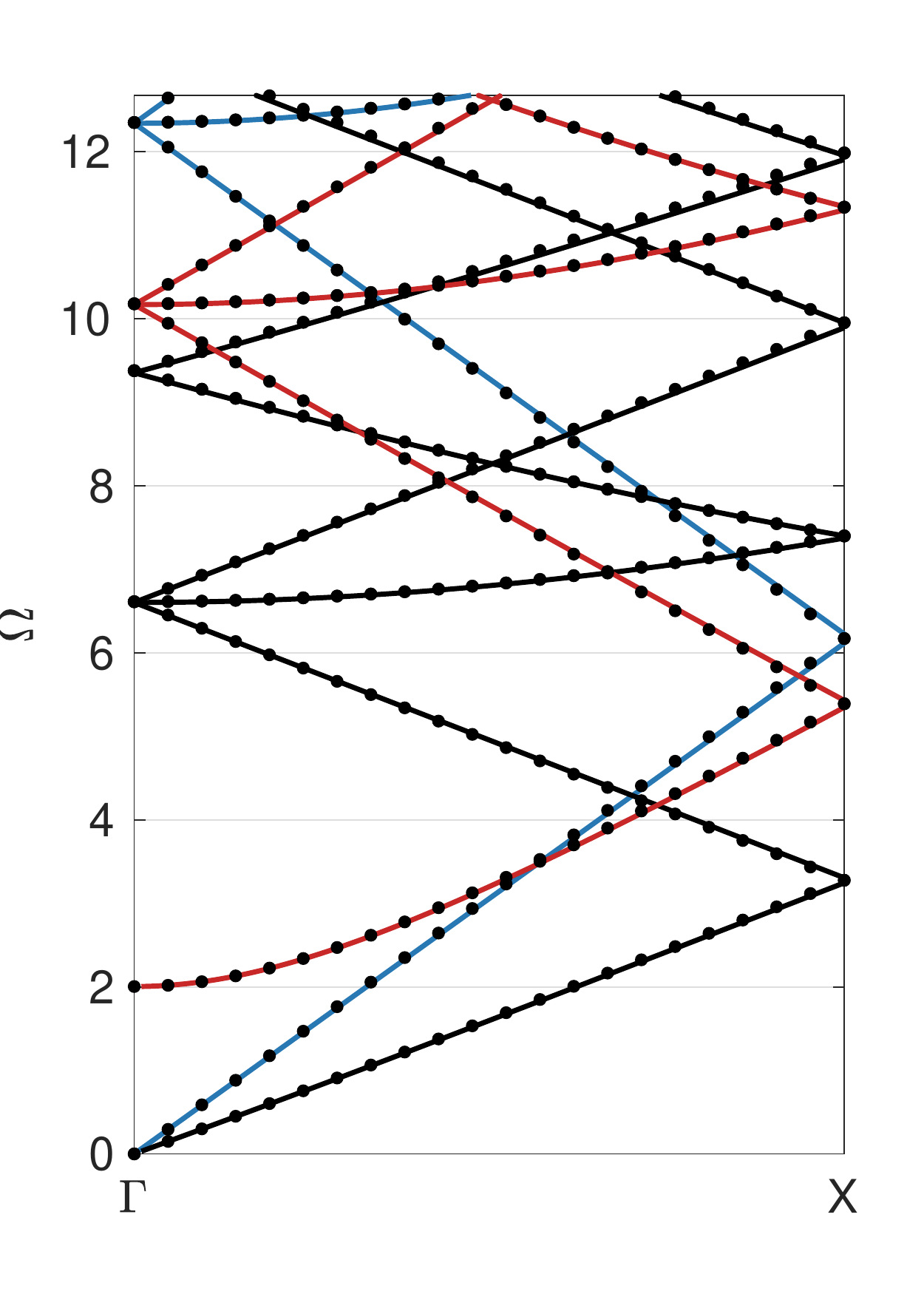}
\label{fig:homocoss}} &
\subfloat{\includegraphics[width=0.1\textwidth]{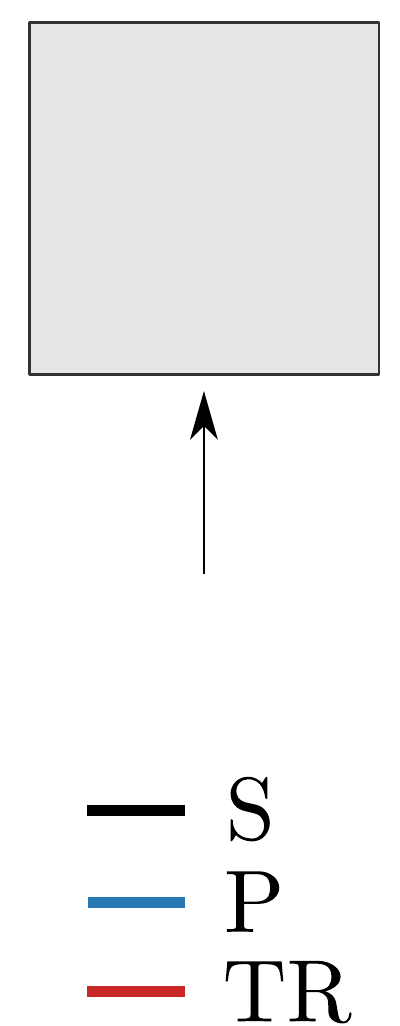}}
\end{tabular}
\caption{Dispersion relations for a homogeneous cell made with material 1 under different kinematic models. Solid line: analytic solution. Dotted line: numerical solution. For all cases $\mathbf{k} = \langle 0,\kappa_y \rangle$.}
\label{fig:homo}
\end{figure}

\Cref{fig:homo} compares the closed-form and numerical results found with the 
user-element subroutine. The presence of an extra degree of freedom with 
respect to the previous one in each model is evidenced by an increasing number 
of dispersion branches in the figure. In the $SH$ model there is a single 
branch corresponding to the horizontally polarized shear wave. The model for 
the in-plane material cell has an additional branch corresponding to the 
longitudinal $P$ mode and with a larger phase speed. Finally, examination of 
the results for the micropolar cell not only show an expected third branch 
corresponding to the rotational wave but they also reveal a dispersive shear 
wave mode. The microrotational wave, denoted as the TR-mode appears after the 
cut--off frequency $\Omega = 2$. This limited existence of the microrotational 
model implies that there is a (microrotational) band-gap in the range $0 < 
\Omega < 2$. This is interesting considering the fact that the material is 
homogeneous. This can be understood like a micro-structural effect which is 
inherent to the kinematics of the continuous micropolar model and that is 
triggered at wavelengths $\lambda>c_T/\Omega_0$, where $\Omega_0$ is the 
cut-off frequency. As a final observation it is worth mentioning that the 
comparison between numerical and the analytical results show a good agreement 
in every considered kinematic model, i.e., out-of-plane, in-plane and 
micropolar. Moreover, the maximum observed relative error between numerical and 
analytical results which occur near near $\Omega = 12$ is less than 0.5\% which 
shows that the subroutine is suitable for computing dispersion relationships 
under different kinematic assumptions.

\subsection{Test of generality about unit cell geometry and materials}

\begin{figure}
  \centering
  \includegraphics[width=3.5 in]{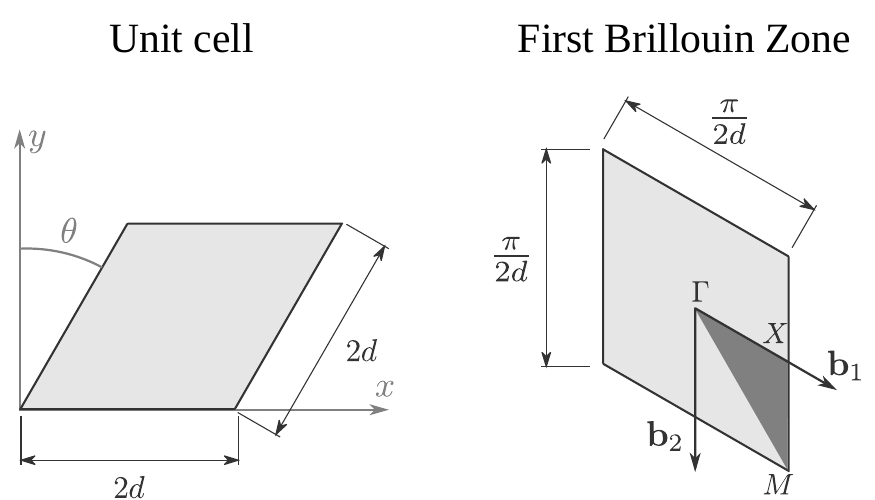}
  \caption{Unit cell and first Brillouin zone for a skewed unit cell.}
  \label{fig:skewed_unit}
\end{figure}

As a second example to show the versatility of the proposed approach we considered unit cells of different geometries and material properties. The first consideration corresponded to a skewed unit cell made of a homogeneous material with properties corresponding to aluminum, as presented in \cref{fig:skewed_unit}. This analysis was later extended to the case of a bi-layer material with properties corresponding to aluminum and brass. These two new cases were analyzed with the in-plane kinematic model and both cases have known closed-form solutions allowing us to compare our results with analytic dispersion curves.

The lattice vectors (\(\mathbf{a}_i\)) and reciprocal lattice vectors  (\(\mathbf{b}_i\)) for the skewed unit cell are \citep{kittel1976introduction} 

\begin{align*}
  &\mathbf{a}_1 = 2d (1, 0)\, ,\quad \mathbf{a}_2 = 2d (\sin\theta, \cos\theta)\, ,\\
  &\mathbf{b}_1 = \frac{2\pi}{d} (1, -\tan\theta)\, ,\quad \mathbf{b}_2 = \frac{2\pi}{d} (0, -\sec\theta)\, ,\\
\end{align*}

while the wavenumbers for the first Brillouin zone read:
\begin{equation}
  k_{m,n} = \sqrt{\left(k_x + \frac{m\pi}{d}\right)^2 + \left(k_y + \frac{m\pi \tan\theta}{d} + \frac{n\pi\sec\theta}{d}\right)^2}\, .
  \label{eq:wavenumber_skew}
\end{equation}

Similarly, the dispersion relations for the bi-layer material under in-plane waves traveling perpendicularly to the layers is given as the following implicit equation \citep{thesis:langlet}:
\[\cos(2dk) = \cos\left(\frac{\omega d}{c_1}\right)\cos\left(\frac{\omega d}{c_2}\right) - \frac{(\rho_1 c_1)^2 + (\rho_2 c_2)^2}{2\rho_1 \rho_2 c_1 c_2}\sin\left(\frac{\omega d}{c_1}\right)\sin\left(\frac{\omega d}{c_2}\right)\, ,\]
where $c_i$ refers to the transverse or longitudinal wave of each layer, and $\rho_i$ refers to the mass density of each layer.

\subsubsection{Numerical results}

\begin{figure}[H]
\centering
\begin{tabular}{m{0.27\textwidth}m{0.27\textwidth}m{0.27\textwidth}m{0.1\textwidth}}
\subfloat[Homogeneous material square unit 
cell]{\includegraphics[width=0.3\textwidth]{img/HOMO_PSV.pdf}
\label{fig:homo_square}} & 
\subfloat[Homogeneous material skewed unit cell]{\includegraphics[width=0.3\textwidth]{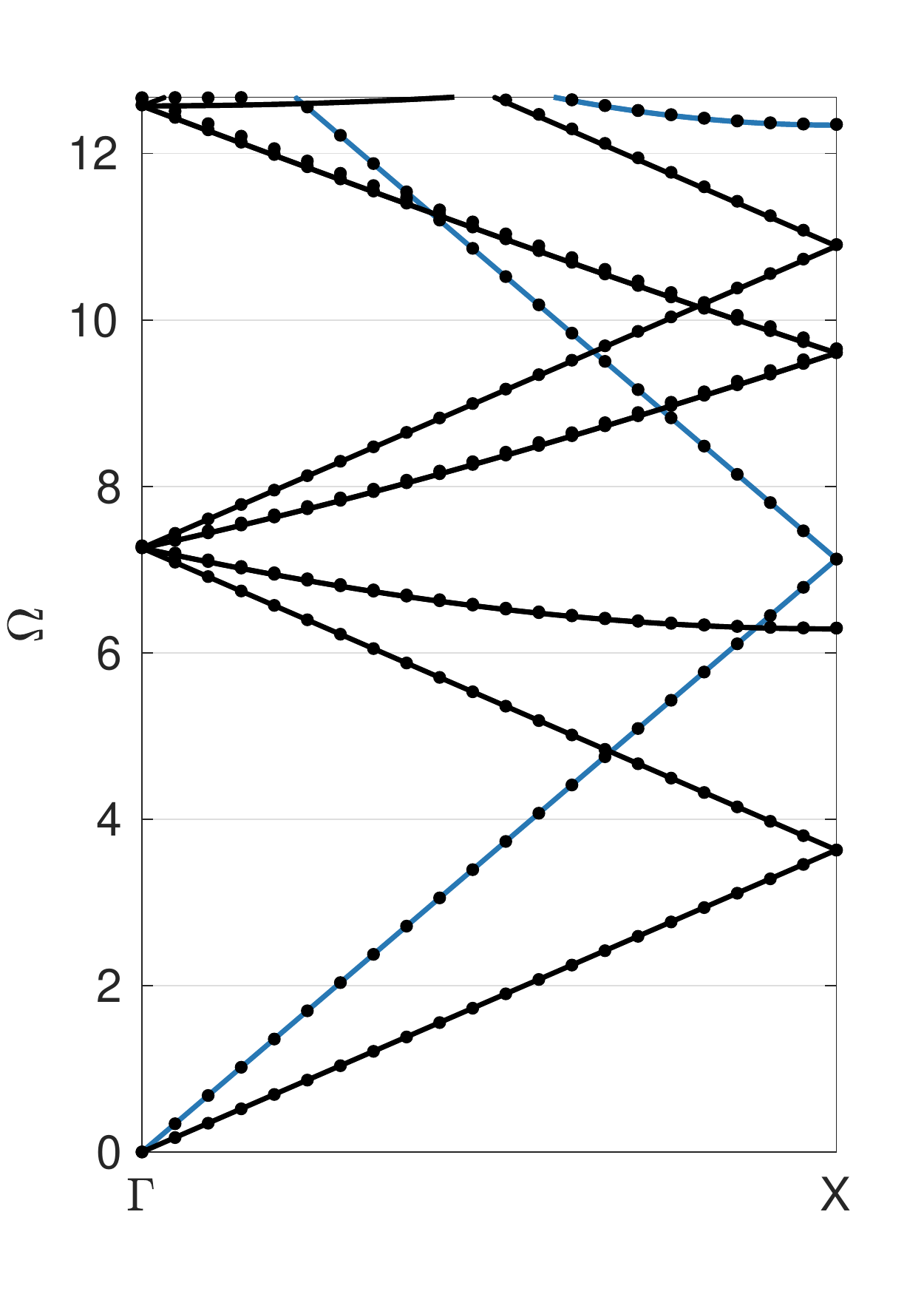}
\label{fig:homo_skewed}} &
\subfloat[Bilayer material unit cell]{\includegraphics[width=0.3\textwidth]{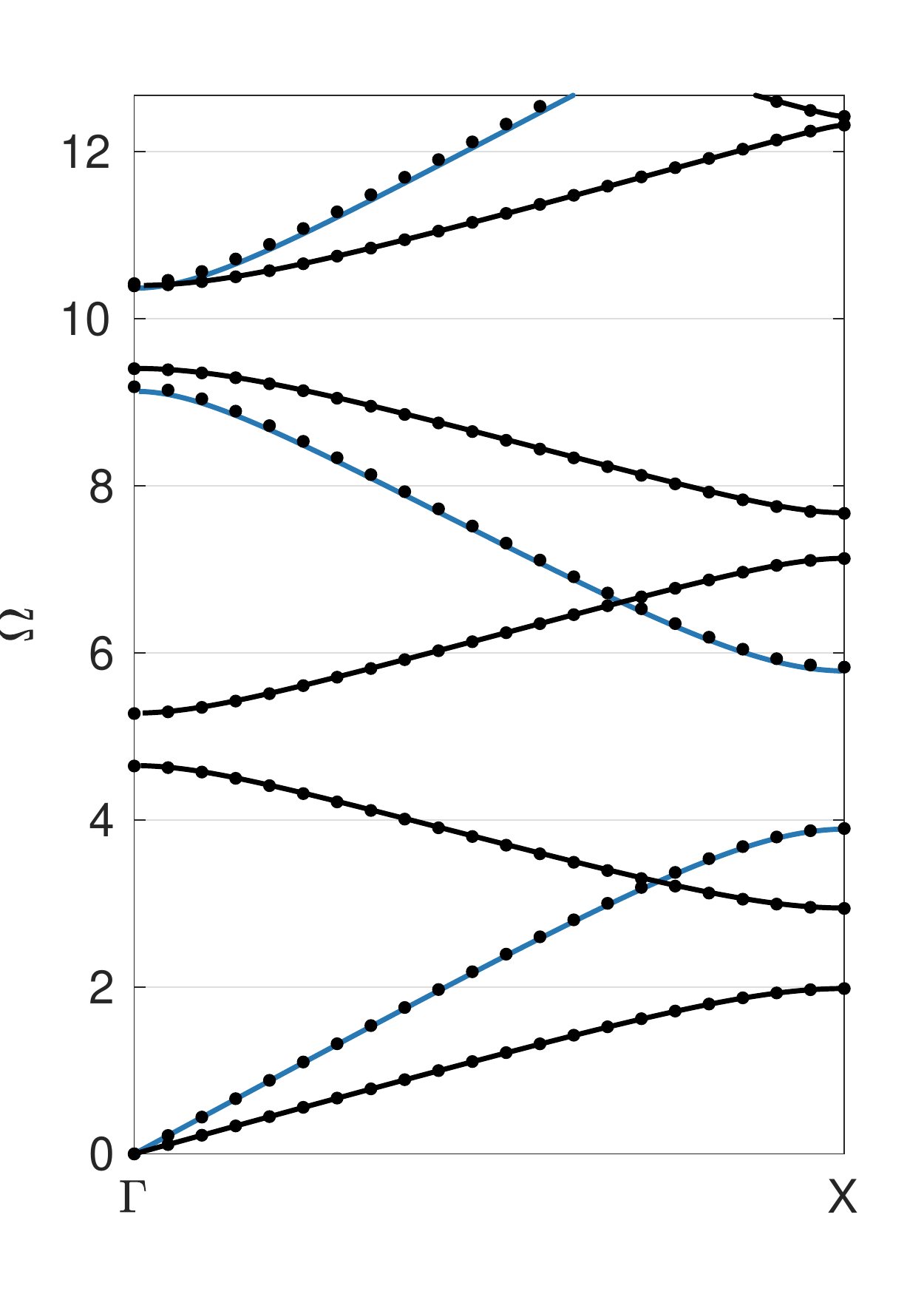}
\label{fig:bilayer}} &
\subfloat{\includegraphics[width=0.1\textwidth]{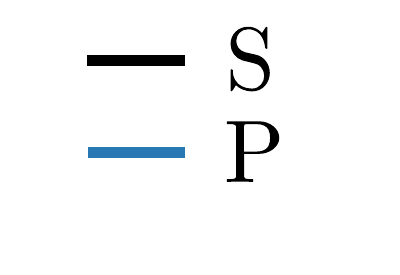}}
\end{tabular}
\caption{Dispersion relations for different cases with known analytical solution under in-plane waves kinematic model. Solid line: analytic solution. Dotted line: numerical solution. For (a) and (c) cases: $\mathbf{k} = \langle 0,\kappa_y \rangle$. For (b) case: $\Gamma X$ direction shown in \cref{fig:skewed_unit}.}
\label{fig:analytical}
\end{figure}

\Cref{fig:homo_square,fig:homo_skewed} show the results for a 2D homogeneous 
material under plane strain idealization computed with a square and skewed unit 
cell respectively. The results from the numerical implementation are in good 
agreement with those predicted by the closed-form dispersion relationships. 
Similarly, \cref{fig:bilayer} shows the results for the bi-layer material with 
vertical wave incidence. In this case the material properties are those of 
aluminum and brass. From the dispersion diagrams It is observed that in the low 
frequency regime ($\Omega < 1$), the bi-layered material behaves as a 
homogeneous material with a linear group velocity. This effective velocity is 
an average between the wave propagation velocities of aluminum and brass. The 
results show once again a very good agreement in comparison to the analytic 
solution. In all cases shown in \cref{fig:analytical} we found a relative error 
under $0.5\%$ near $\Omega = 12$.

\subsection{Verification against external numerical results}

\begin{figure}[H]
\centering
\begin{tabular}{m{0.1\textwidth}m{0.27\textwidth}m{0.27\textwidth}m{0.1\textwidth}}
\includegraphics[width=0.1\textwidth]{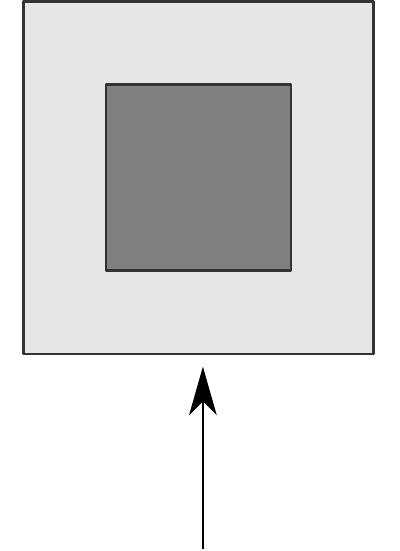}&
\subfloat[Squared inclusion]{\includegraphics[width=0.3\textwidth]{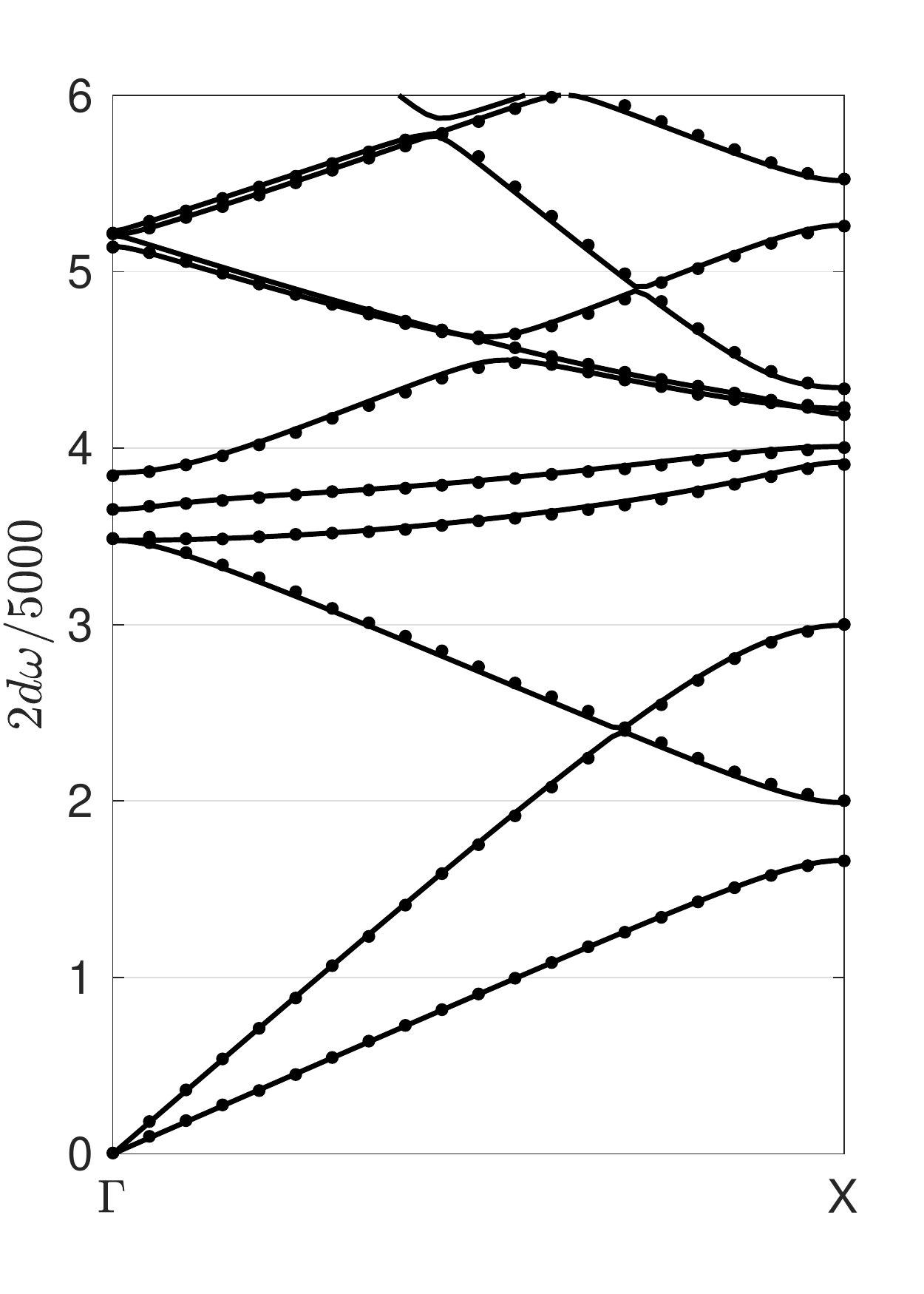}
\label{fig:cuad_lan}}&
\subfloat[Circular pore]{\includegraphics[width=0.3\textwidth]{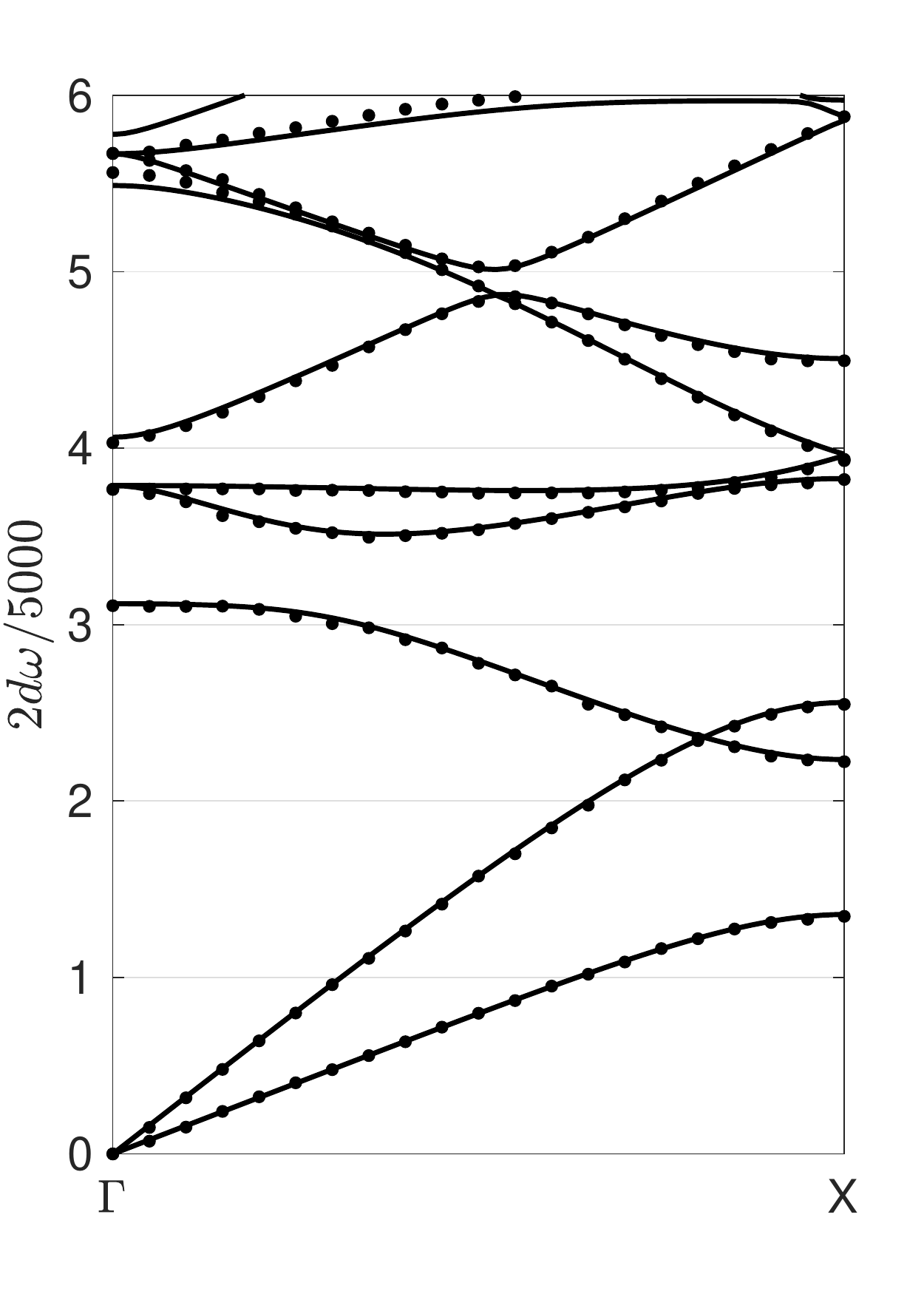}
\label{fig:cir_lan}}&
\includegraphics[width=0.1\textwidth]{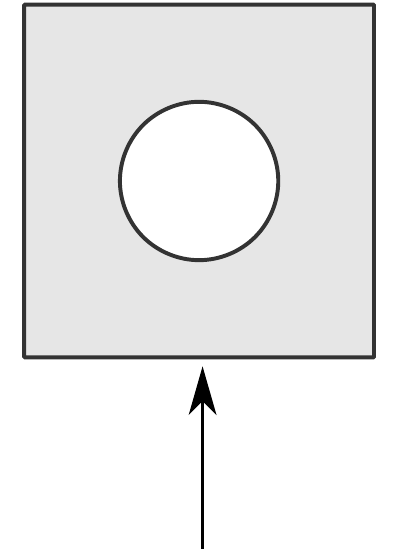}
\end{tabular}
\caption{Dispersion relations for a squared inclusion and a circular pore unit cells. Solid line: Current strategy solution. Dotted line: results from \cite{thesis:langlet}. For both cases $\mathbf{k} = \langle 0,\kappa_y \rangle$.}
\label{fig:lang}
\end{figure}

As a final verification, we compared our numerical solutions with the results reported by \cite{thesis:langlet} corresponding to microstructures in the form of a squared inclusion in a homogeneous matrix, and of a circular pore in a squared unit cell. In the first case, the inclusion is made of brass with an aluminum-based matrix. The ratio between the characteristic dimensions of the inclusion and the cell in each case corresponds to $a_s/2d = 1/3$ $a_p/2d = 1/2$. The comparison between our results and those obtained by the independent numerical implementation reported in \cite{thesis:langlet} shown in \cref{fig:lang} are in good agreement. 

\subsection{Additional results}
\begin{figure}[H]
  \centering
  \includegraphics[width=5 in]{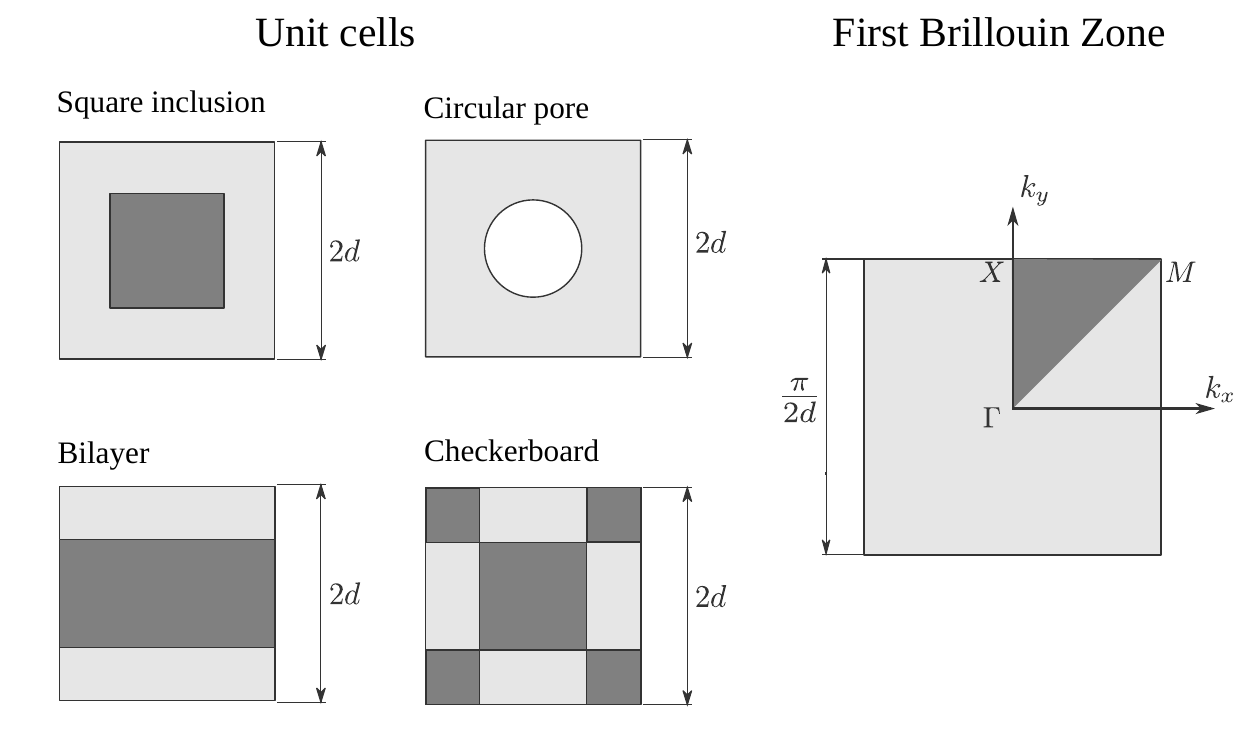}
  \caption{(Left) Unit cells used in the analyses. Light gray: Material 1, dark gray: Material 2. (Right) Illustration of first Brillouin zone and the irreducible Brillouin zone for the considered analyses. The unit cells were selected to have the same irreducible Brillouin zone.}
\label{fig:cells}
\end{figure}

As a final test of the capabilities in the user-element subroutine we combined the three different kinematic models discussed previously with multiple microstructural configurations corresponding to bilayer material, circular pore, square and checkerboard inclusion. The unit cells for these materials together with the first Brillouin zone are depicted in \cref{fig:cells}. Due to the symmetry of the considered cases all results are presented along this irreducible Brillouin zone. 

\begin{figure}[H]
\centering
\begin{tabular}{m{0.27\textwidth}m{0.27\textwidth}m{0.27\textwidth}m{0.1\textwidth}}
\subfloat[Out-of-plane]{\includegraphics[width=0.3\textwidth]{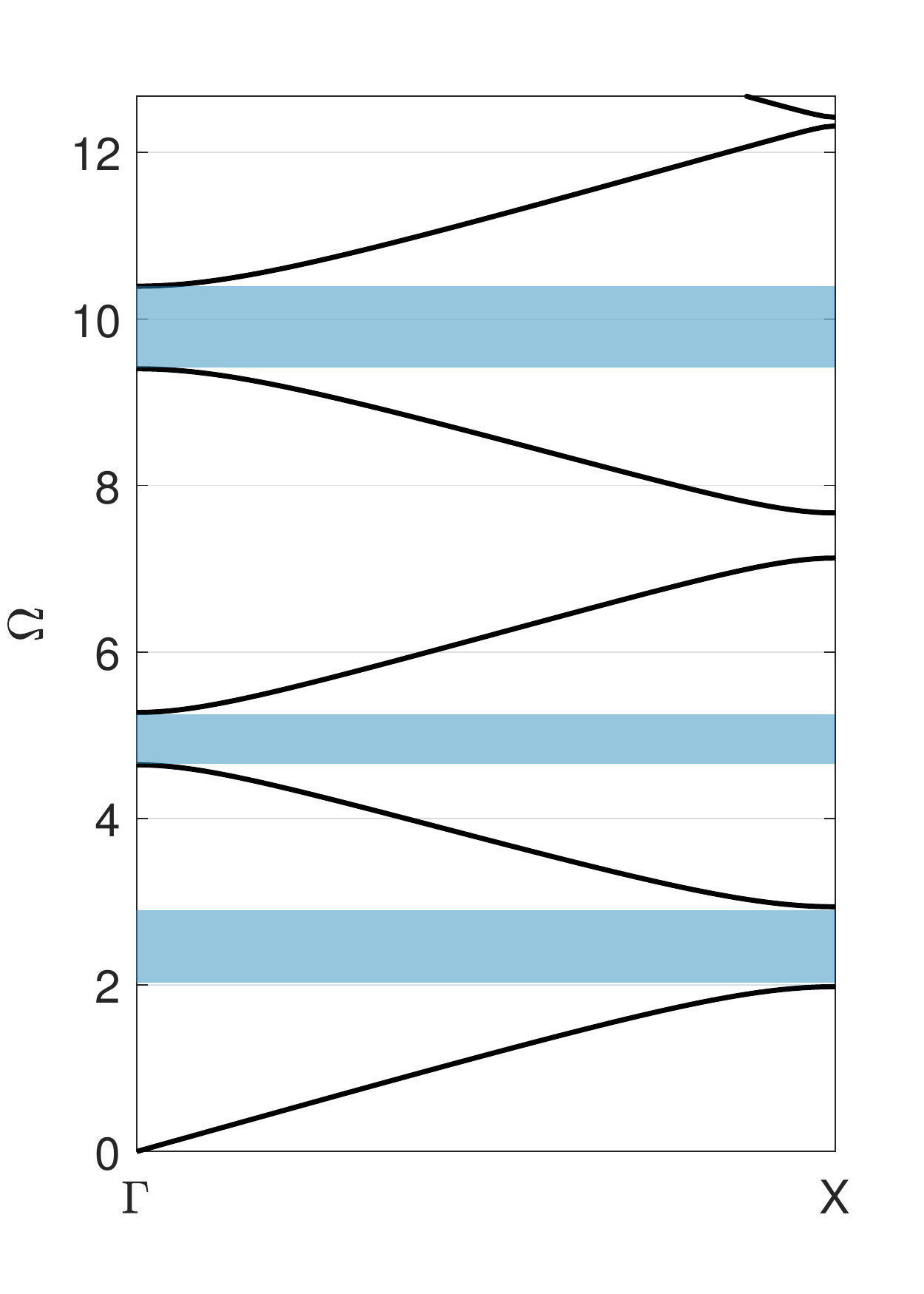}
\label{fig:bicapash}} &
\subfloat[In-plane]{\includegraphics[width=0.3\textwidth]{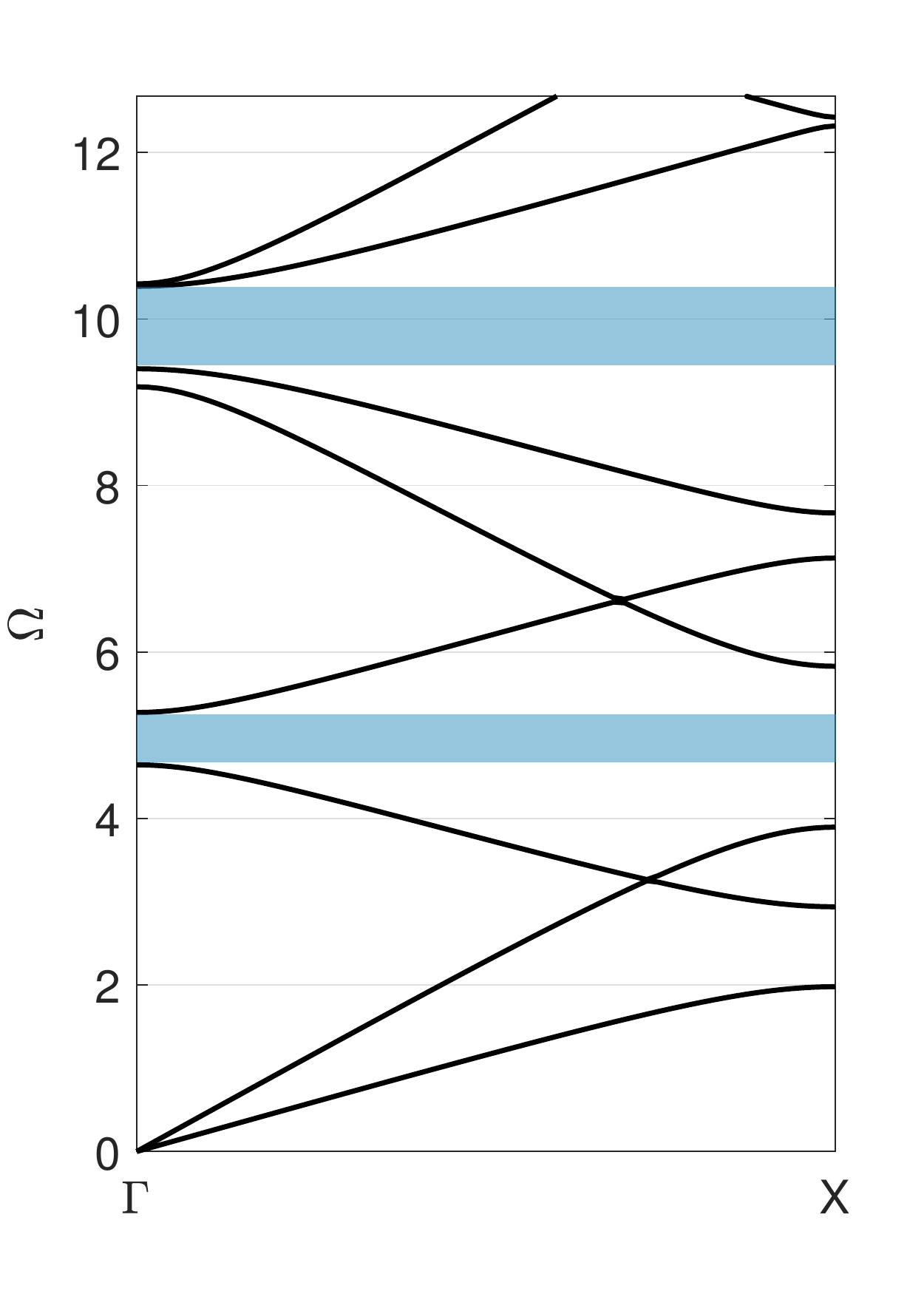}
\label{fig:bicapapsv}} &
\subfloat[Micropolar]{\includegraphics[width=0.3\textwidth]{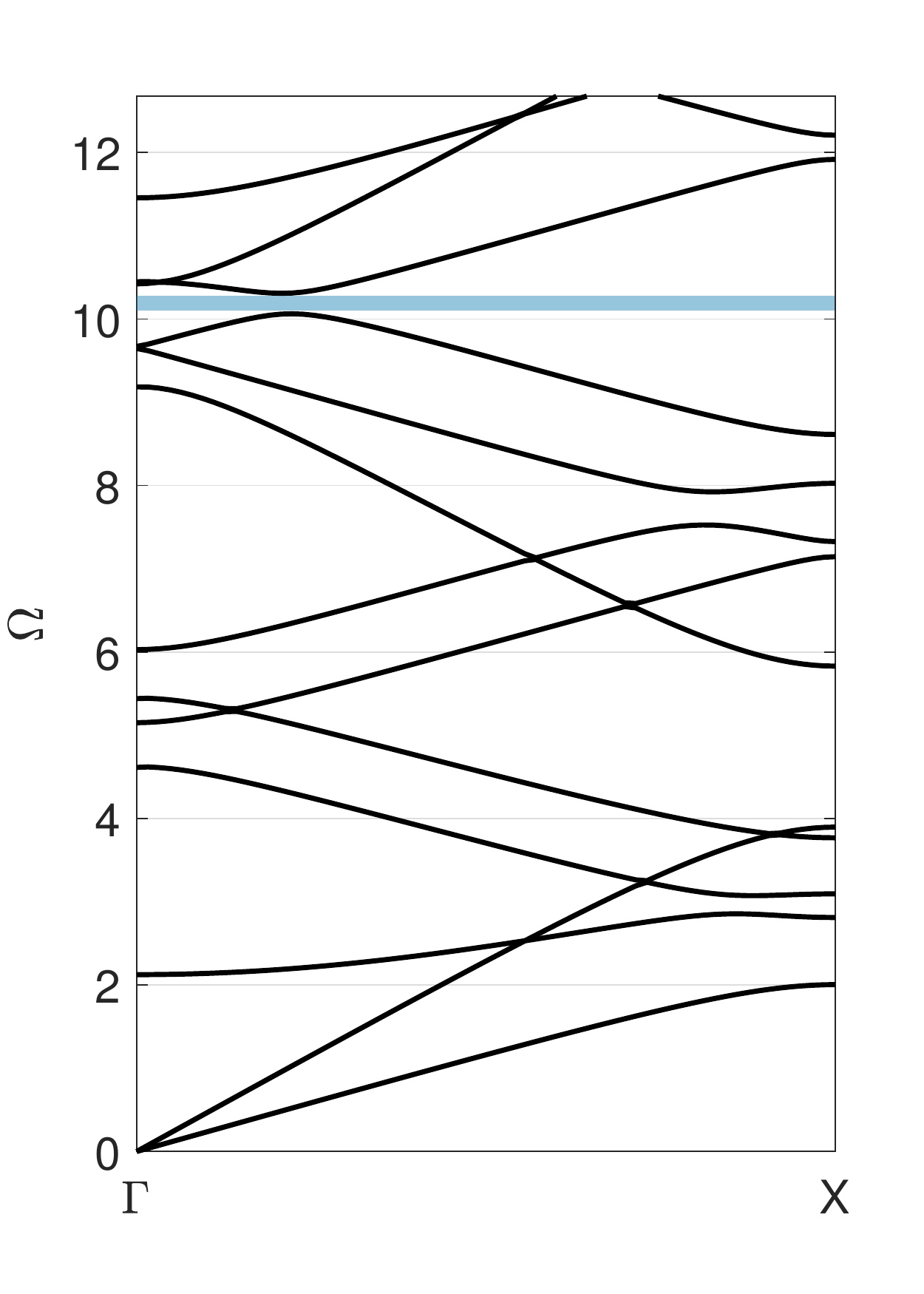}
\label{fig:bicapacoss}} &
\includegraphics[width=0.1\textwidth]{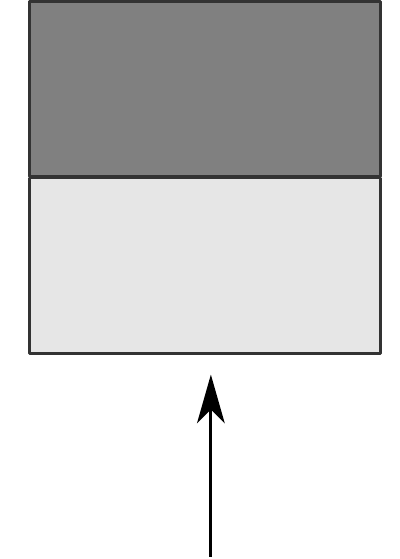}
\end{tabular}
\caption{Dispersion relations for a bilayer material made with materials 1 and 2 under different kinematic models. Each layer has a thickness $t = d$. The analysis was made considering only the direction in which the material is periodic.}
\label{fig:bicapa}
\end{figure}

\Cref{fig:bicapa} shows the results for a bilayer material with base properties as per \cref{tab:prop}. There are three $S$ wave band gaps (shown by the shaded rectangle) in the $SH$ model all of them occurring along the vertical $\Gamma X$ direction which is precisely the direction of periodicity  of the microstructure. The lower frequency bandgaps are interrupted by the $P$ wave modes once we consider the in-plane behavior. This pattern of elimination of bandgaps as we consider additional degrees of freedom is also observed in the micropolar case shown in \cref{fig:bicapacoss}

\begin{figure}[H]
\centering
\begin{tabular}{m{0.27\textwidth}m{0.27\textwidth}m{0.27\textwidth}m{0.1\textwidth}}
\captionsetup[subfigure]{labelformat=empty}
\subfloat{\includegraphics[width=0.3\textwidth]{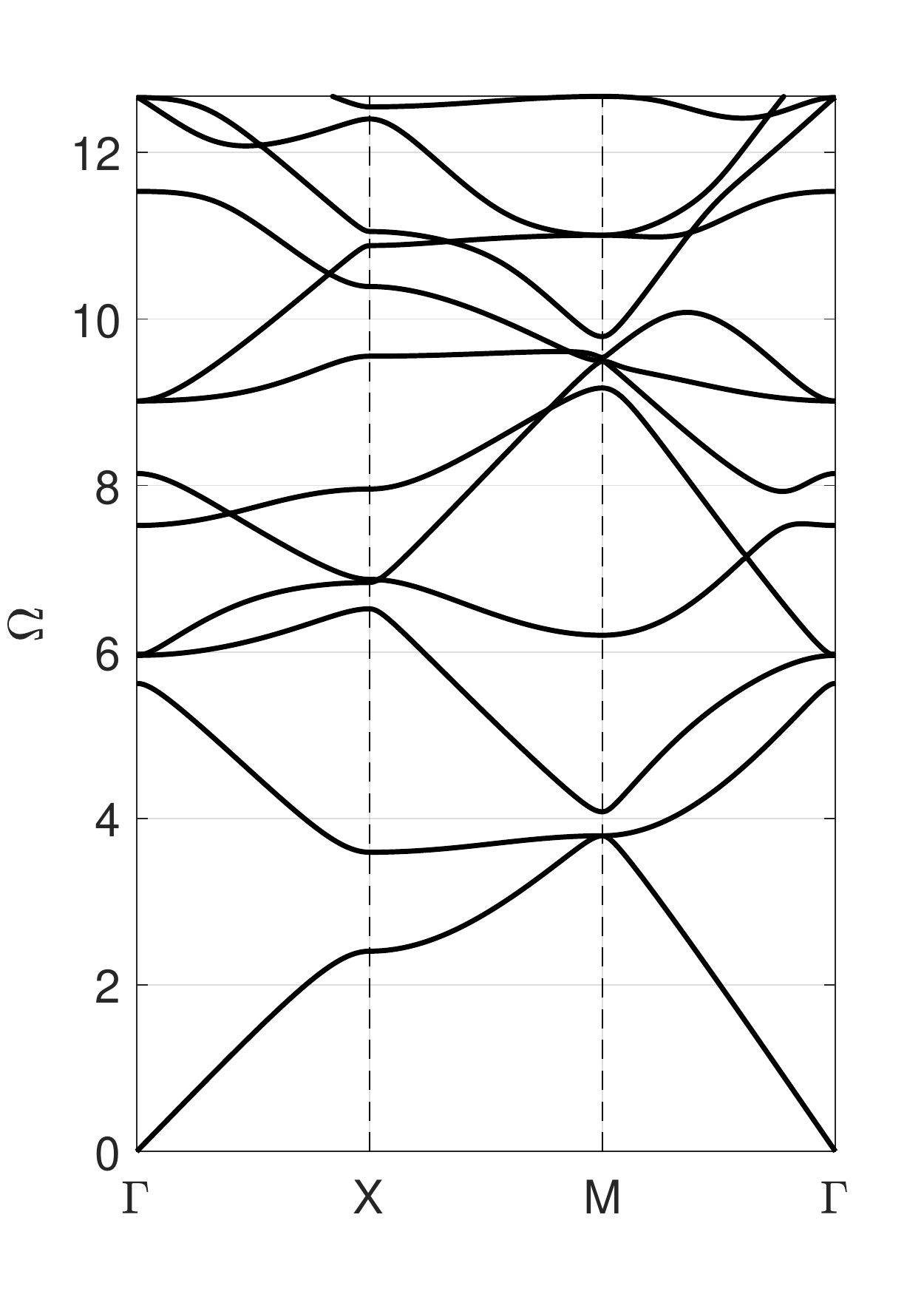}} &
\subfloat{\includegraphics[width=0.3\textwidth]{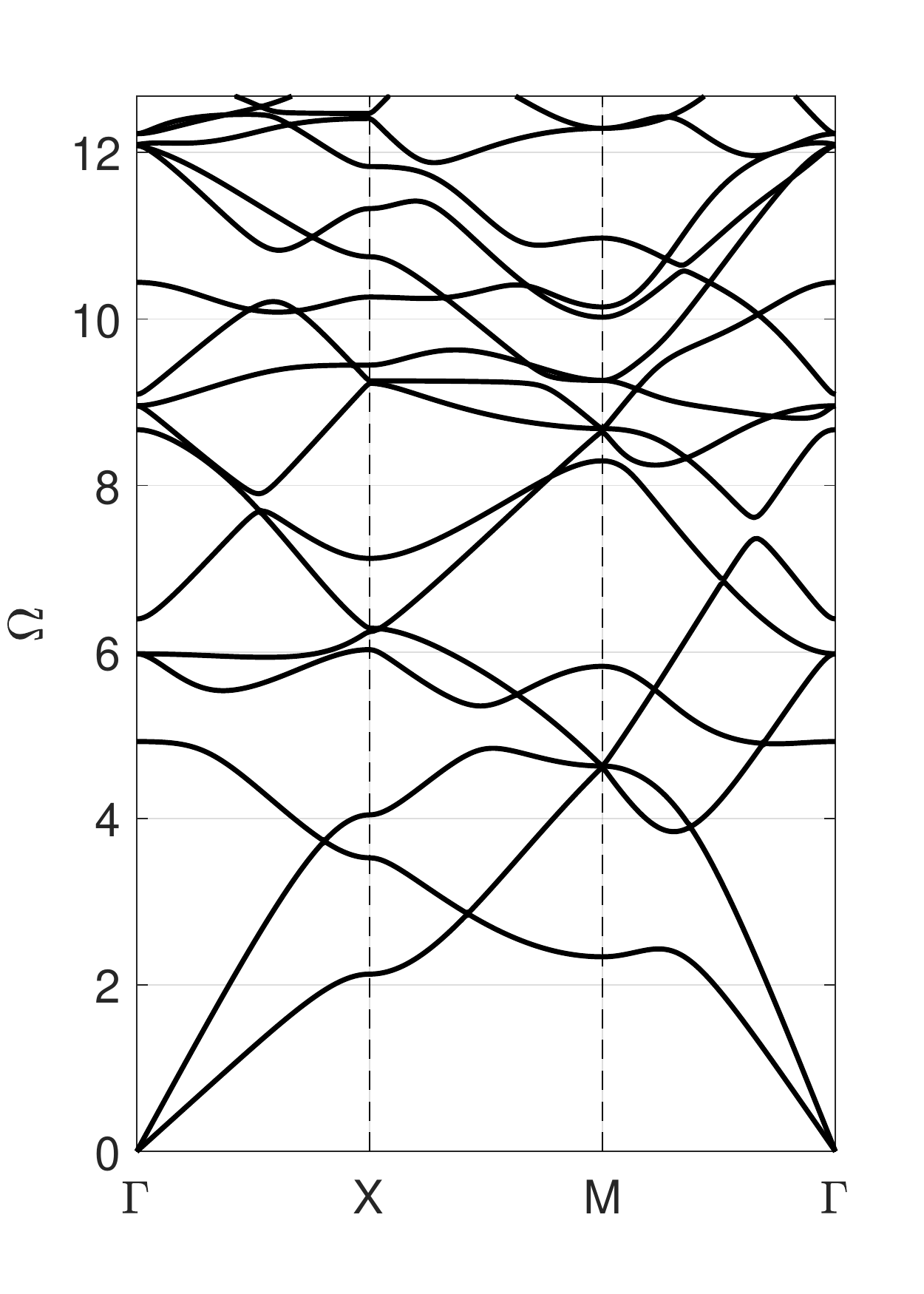}}&
\subfloat{\includegraphics[width=0.3\textwidth]{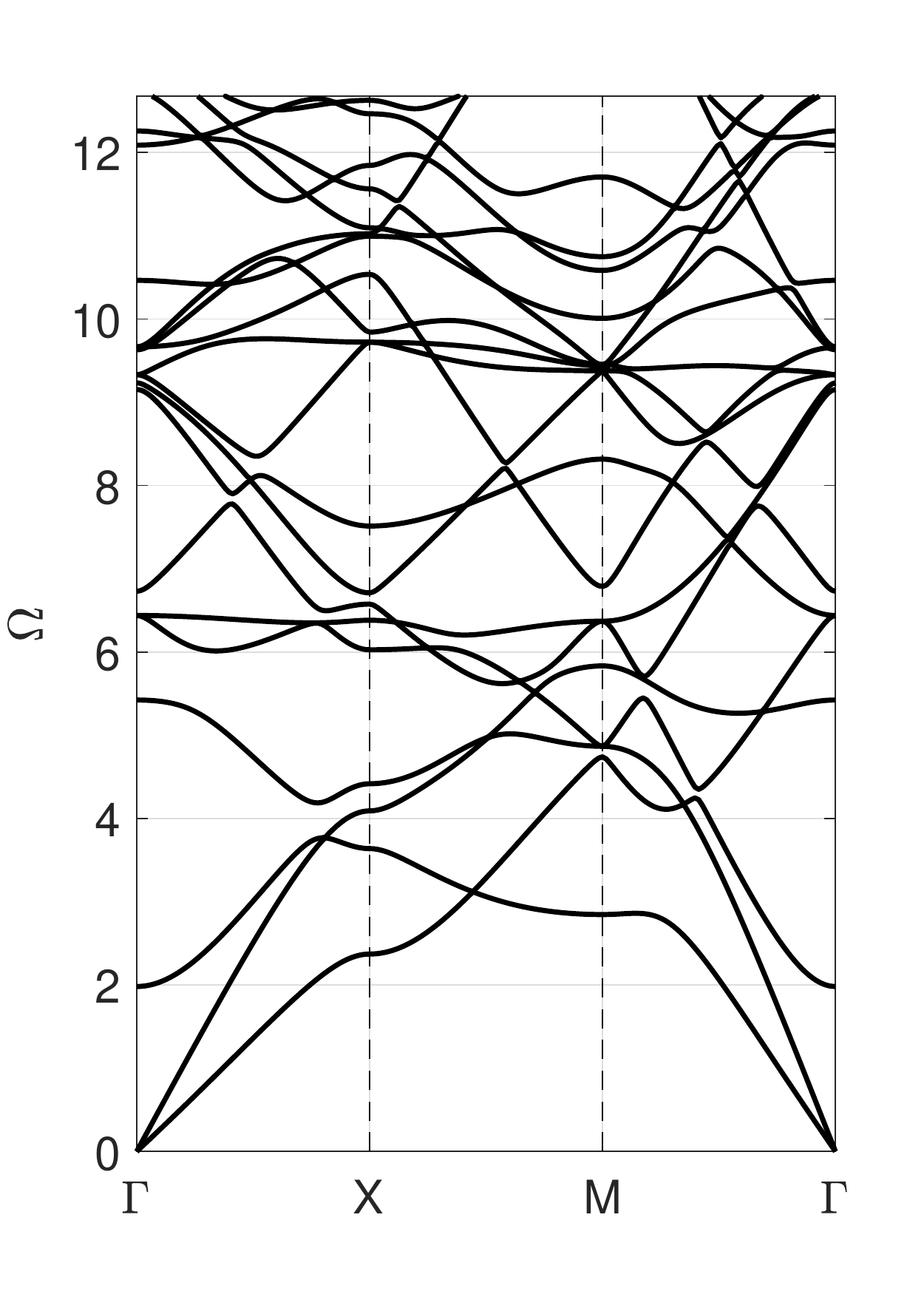}}& 
\includegraphics[width=0.1\textwidth]{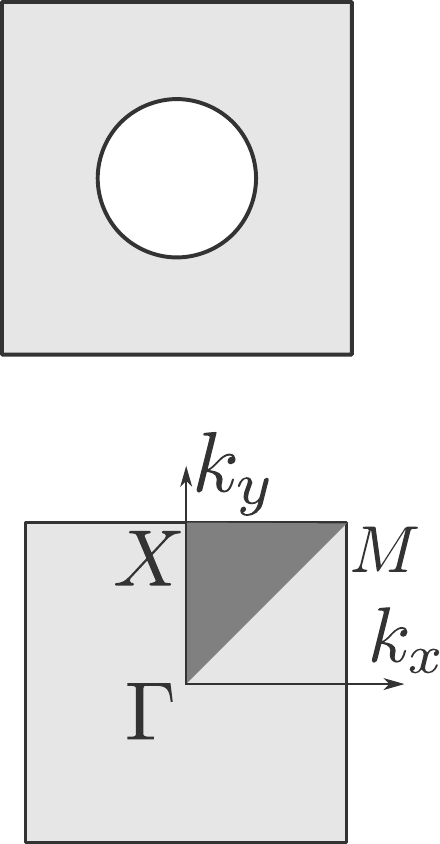} \\
\subfloat{\includegraphics[width=0.3\textwidth]{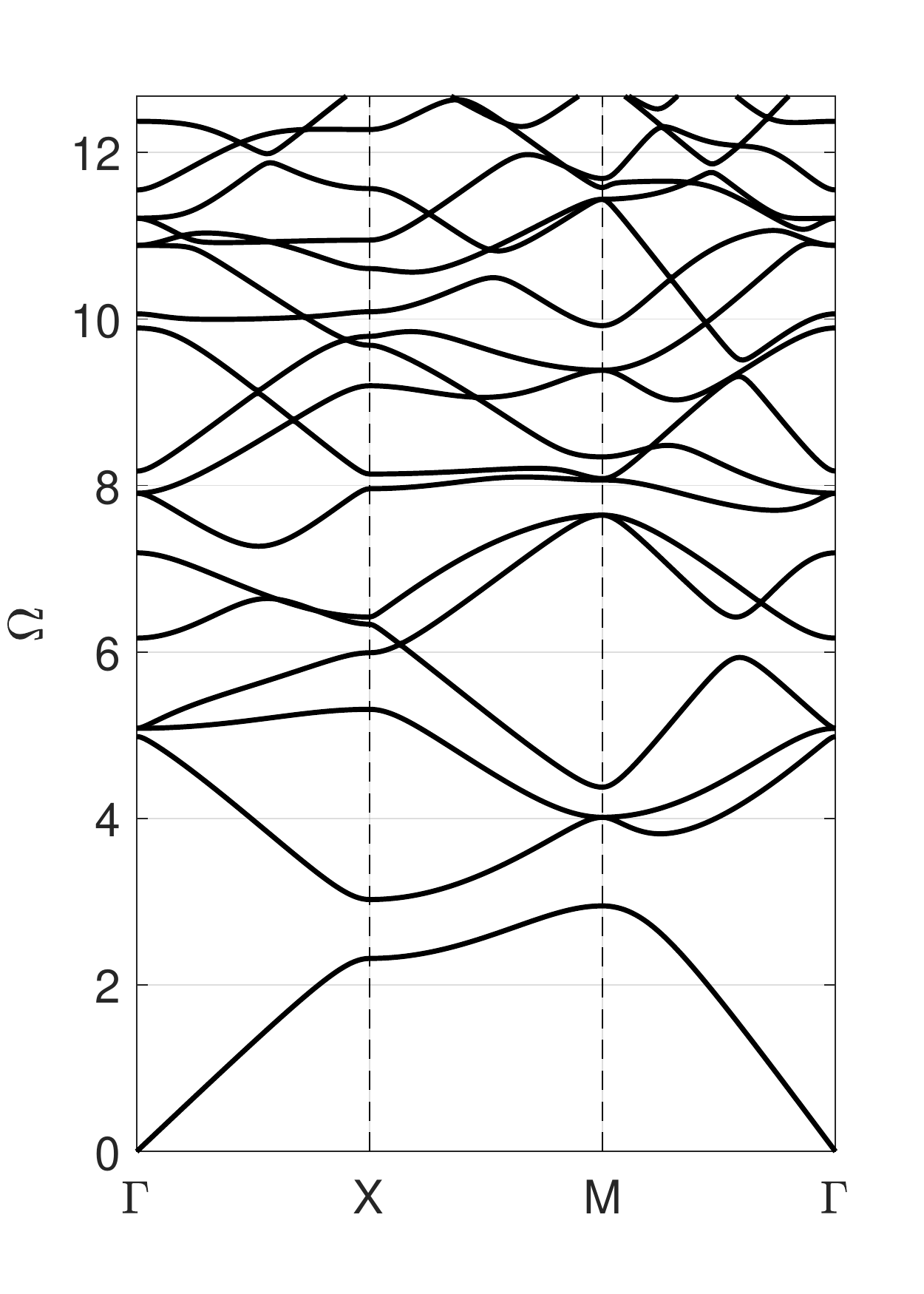}}&
\subfloat{\includegraphics[width=0.3\textwidth]{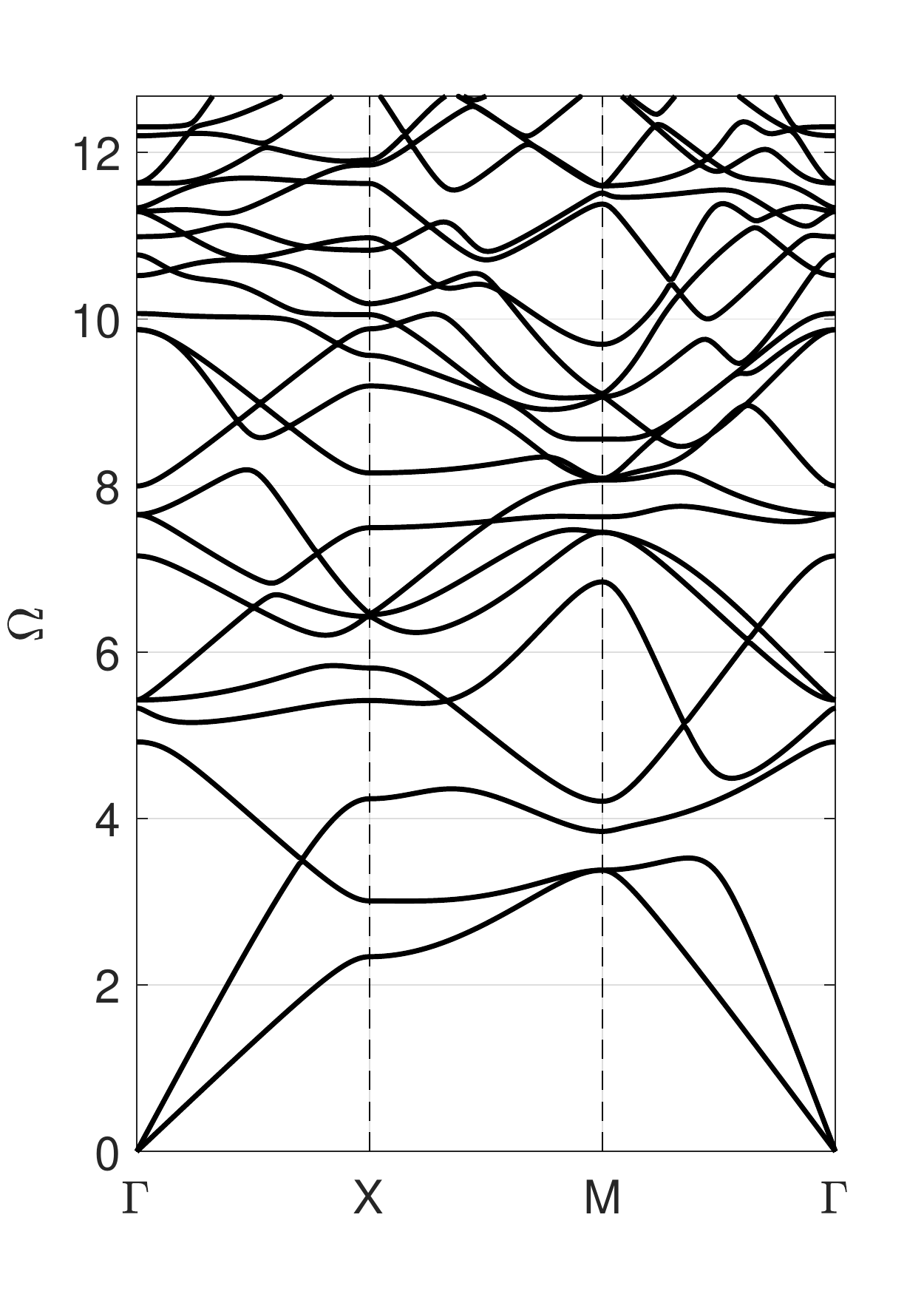}}&
\subfloat{\includegraphics[width=0.3\textwidth]{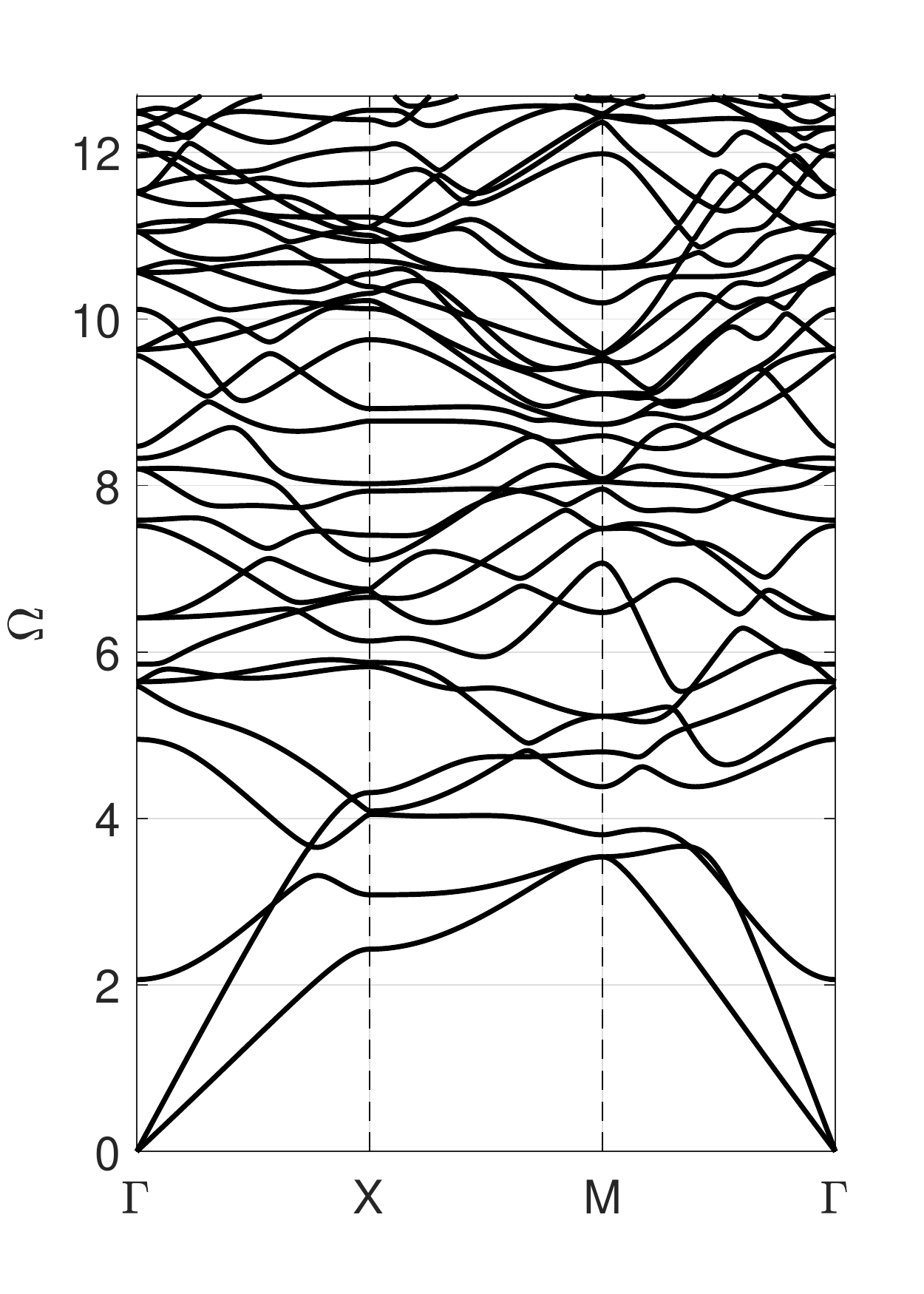}}&
\includegraphics[width=0.1\textwidth]{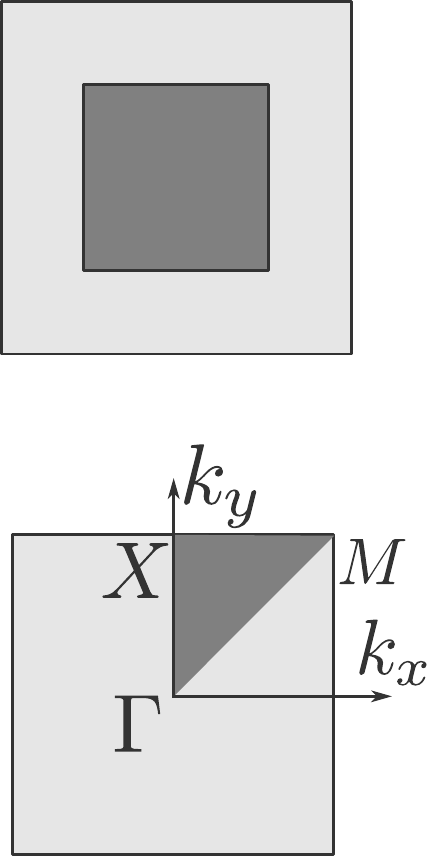} \\ 
\subfloat[Out-of-plane]{\includegraphics[width=0.3\textwidth]{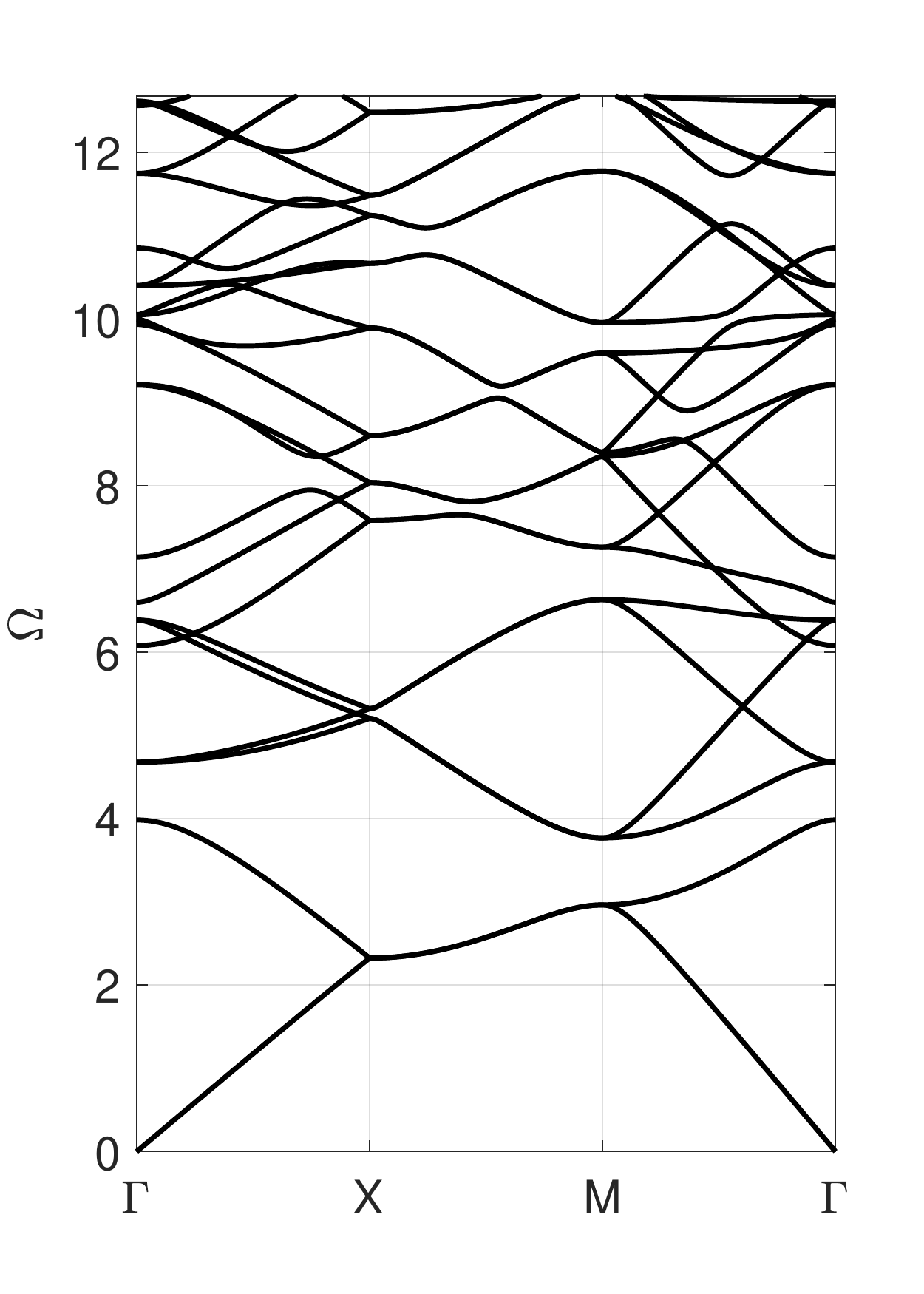}}&
\subfloat[a][In-plane]{\includegraphics[width=0.3\textwidth]{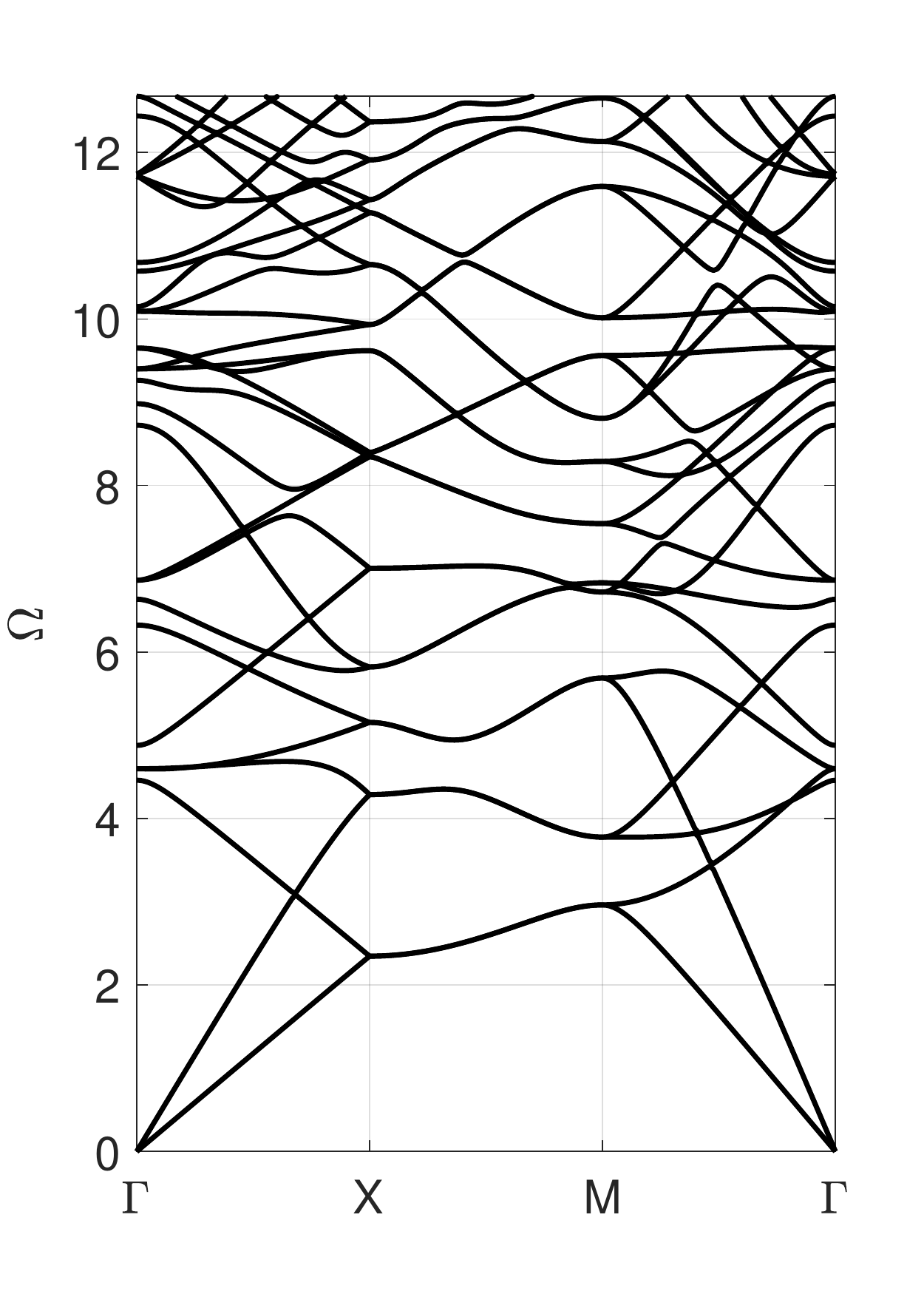}}&
\subfloat[Micropolar]{\includegraphics[width=0.3\textwidth]{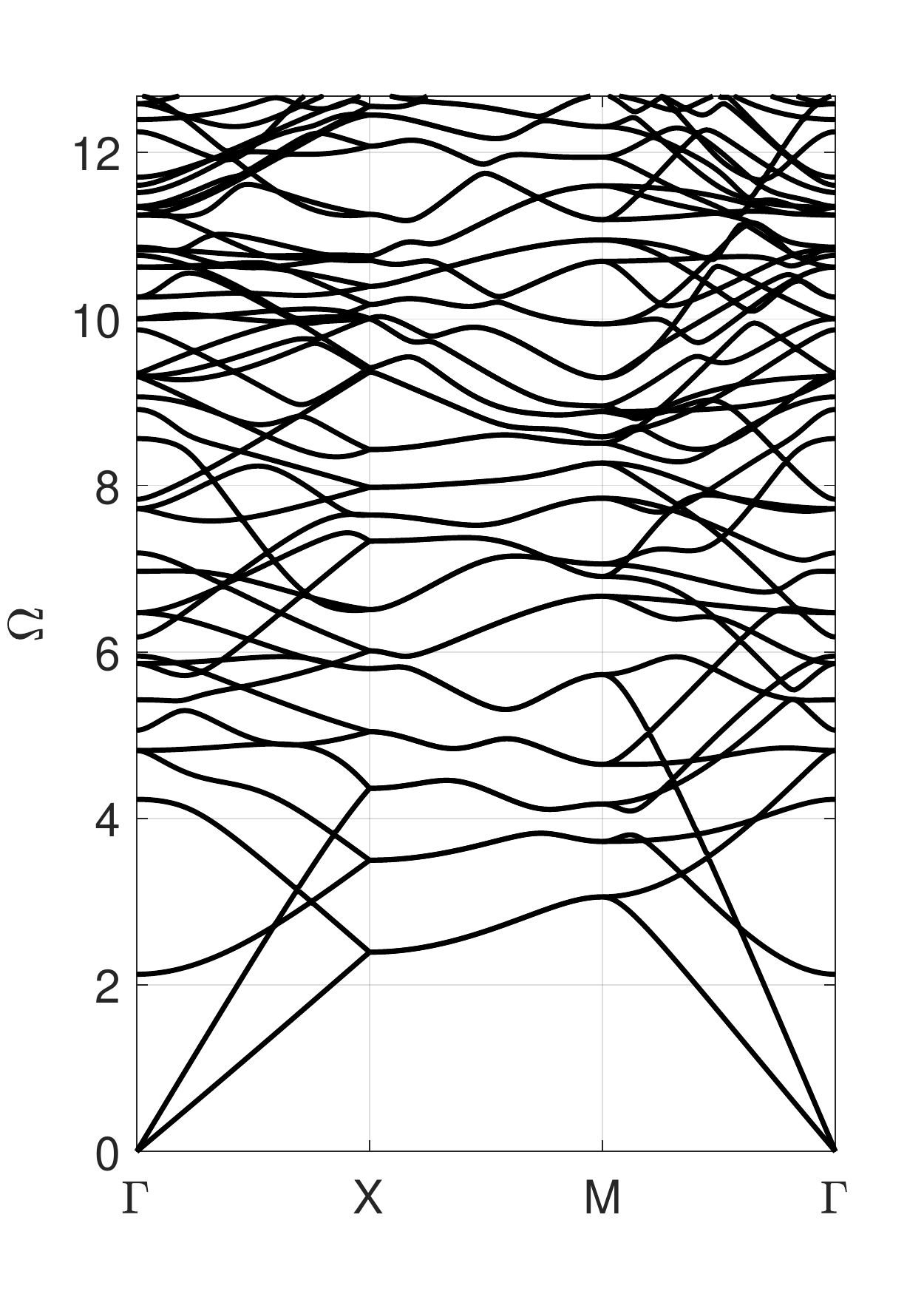}}&
\includegraphics[width=0.1\textwidth]{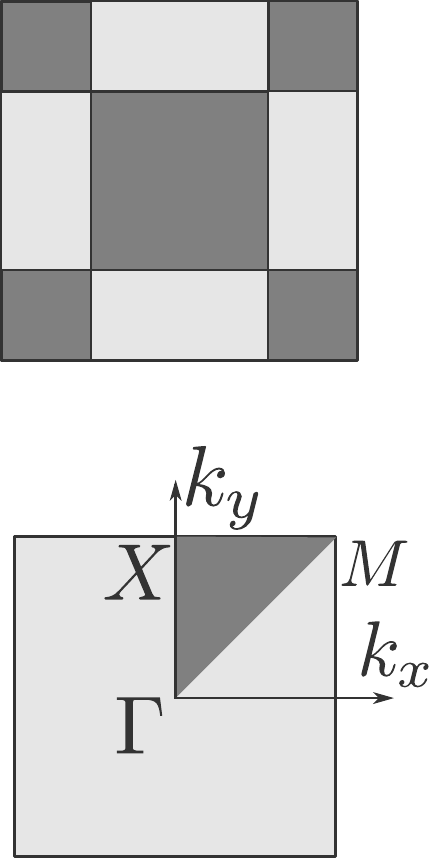}
\end{tabular}
\caption{Dispersion relations for different unit cells under different kinematic models. First row: Circular pore. Second row: Square inclusion. Third row: checkerboard. Columns 1 to 3 correspond to out-of-plane, in-plane and micropolar kinematic models respectively.}
\label{fig:todas}
\end{figure}

Similarly, in \cref{fig:todas} we present the results for the circular pore, 
and square and checkerboard inclusions. In all unit cells, the out-of-plane 
case exhibits partial band gaps along the $\Gamma X$ direction. As observed 
previously these gaps are progressively eliminated as we introduce additional 
degrees of freedom into the model. For the micropolar model the rotational wave 
has a bandgap in the range $0 < \Omega < 2$, but for $\Omega < 2$, the 
responses with the micropolar and classical model are equivalent. The results 
for this last set of microstructures and kinematic assumptions are physically 
consistent as indicated by the additional branches obtained as we increased the 
number of DOFs. Also in the low frequency regime (under $\Omega = 1$), the 
group speed has a linear behavior as expected. This implies that for waves with 
wavelength values $\lambda >> 2d$ (i.e., propagating at low frequencies) any
microstructure will be blind to the propagating disturbance.

\section*{Conclusions}
We have presented a novel and easy-to-implement computational alternative to 
obtain dispersion relationships in periodic phononic crystals using commercial 
(or already existing) finite element codes. The proposed approach has two basic 
appealing features: first, it avoids manipulating the global arrays, which is a 
condition present in most codes; and second Bloch periodic boundary conditions 
are imposed at the element level thus allowing for the consideration of 
problems belonging to different physical contexts. The approach is based on the 
creation of two assembly operators for those elements $m$ subjected to Bloch 
boundary conditions, namely a $C_m^c$ operator, which is used to retrieve 
element nodal coordinates and other relevant geometric information and an 
operator termed $C_m^a$ which assembles the element into the global arrays 
considering the proper boundary conditions at the onset. At the same time, the 
generality in the proposed technique makes the implementation to 2D and 3D 
problems equally easy. Moreover, the proposed strategy can be directly 
incorporated into existing user element subroutines just through subtle 
changes. Once implemented, the technique is straightforward to use since the 
user just needs to input the coordinate-connectivity operator $C_m^c$ and the 
assembly-connectivity operator $C_m^a$ for each element in the mesh. On the 
other hand, the complex-valued nature of the global arrays arising as a 
consequence of Bloch periodic conditions is dealt with using a duplicate-mesh 
approach reported in the literature. The method and its implementation was 
verified against closed-form solutions for a homogeneous and a bi-layer 
material under different kinematic assumptions and geometries for the material 
cell. We also conducted additional verification exercises using numerical 
results reported by \cite{thesis:langlet}. Finally, a detailed description to 
implement the presented strategy is provided in the supplementary material of 
this work in terms of a compiled version of FEAPpv with user element 
subroutines to compute dispersion relations.

\section*{Acknowledgements}
This work was supported by EAFIT and COLCIENCIAS' Scholarship Program No. 6172.
Preprint of an article published in Journal of Theoretical and Computational 
Acoustics, © \href{https://www.worldscientific.com/worldscinet/jca}{World 
Scientific Publishing Company}.
 
\bibliographystyle{unsrt}
\bibliography{ref.bib}

\end{document}